\documentclass[10pt,journal,compsoc]{IEEEtran}

\ifCLASSOPTIONcompsoc
  \usepackage[nocompress]{cite}
\else
  \usepackage{cite}
\fi

\usepackage{mathptmx}
\usepackage{graphicx}
\usepackage{times}
\usepackage{amsmath}
\usepackage{hyperref}
\usepackage{caption,subcaption}
\usepackage{multirow}
\usepackage{paralist}
\usepackage{listings}

\usepackage{color}
\definecolor{RED}{rgb}{1,0,0}
\definecolor{BLUE}{rgb}{0,0,1}

\newcommand{\sref}[1]{Section~\ref{#1}}
\newcommand{\fig}[1]{Figure~\ref{#1}}

\begin{document}
%
\title{Equalizer 2.0 --\\Convergence of a Parallel Rendering Framework}
%
%
%
%

\author{Stefan~Eilemann\thanks{email: eilemann@gmail.com},
        David~Steiner\thanks{email: steiner@ifi.uzh.ch} \\ %
        and~Renato~Pajarola\thanks{email:pajarola@acm.org},~\IEEEmembership{Senior Member,~IEEE}%
\IEEEcompsocitemizethanks{\IEEEcompsocthanksitem All authors are with the
Visualization and MultiMedia Lab, Department of Informatics, University of
Z\"urich.}}



%
%

\markboth{Journal of \LaTeX\ Class Files,~Vol.~14, No.~8, August~2015}%
{Shell \MakeLowercase{\textit{et al.}}: Bare Demo of IEEEtran.cls for Computer
Society Journals}
%


\IEEEtitleabstractindextext{%
  \begin{abstract}
    Developing complex, real world graphics applications which leverage multiple
    GPUs and computers for interactive 3D rendering tasks is a complex task. It
    requires expertise in distributed systems and parallel rendering in addition
    to the application domain itself. We present a mature parallel rendering
    framework which provides a large set of features, algorithms and system
    integration for a wide range of real-world research and industry
    applications. Using the \textsf{Equalizer} parallel rendering framework, we
    show how a wide set of generic algorithms can be integrated in the framework
    to help application scalability and development in many different domains,
    highlighting how concrete applications benefit from the diverse aspects and
    use cases of \textsf{Equalizer}. We present novel parallel rendering
    algorithms, powerful abstractions for large visualization setups and virtual
    reality, as well as new experimental results for parallel rendering and data
    distribution.
  \end{abstract}

  \begin{IEEEkeywords}
    Parallel Rendering, Scalable Visualization, Cluster Graphics, Immersive
Environments, Display Walls
  \end{IEEEkeywords}
}

\maketitle

\IEEEdisplaynontitleabstractindextext

%
\IEEEpeerreviewmaketitle

\begin{figure*}[ht]\center
  \includegraphics[width=2\columnwidth]{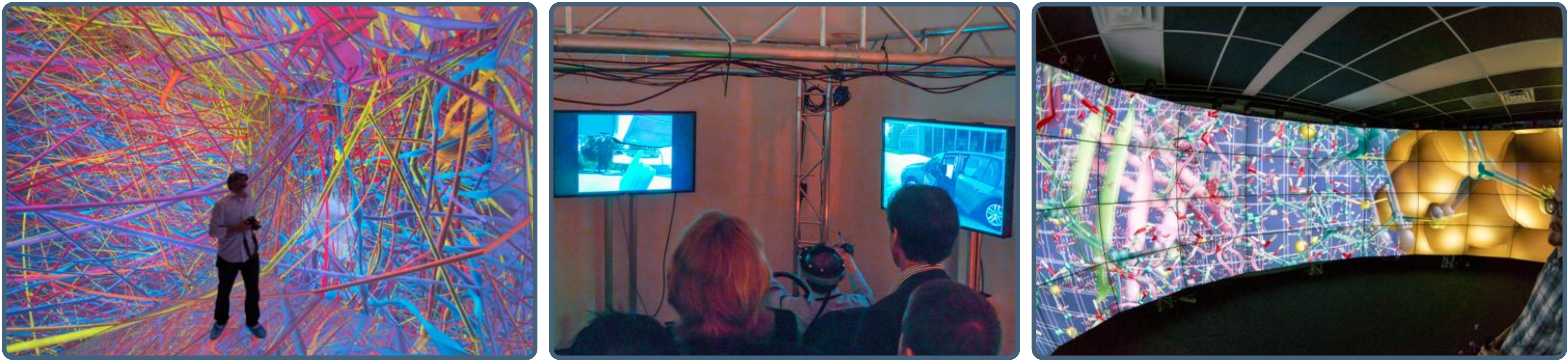} \\
  (a) \hfil \hfil (b) \hfil \hfil (c)
  \vspace{-2mm}
  \caption{Example Equalizer applications: (a) 192 Megapixel CAVE at
    KAUST running RTNeuron, (b) Immersive HMD with external tracked and
    untracked views running RTT DeltaGen for virtual car usability studies,
    and (c) Cave2 running a molecular visualization.}
  \label{FIG_teaser}
\end{figure*}

\IEEEraisesectionheading{\section{Introduction} \label{sec:introduction}}

The Equalizer parallel rendering framework as first presented in~\cite{EMP:09} has
demonstrated its general versatility and the usefulness of its minimally-intrusive
programming design in a variety of applications and projects.
In particular, the integration of large-scale parallel rendering algorithms,
APIs for developing complex distributed applications, and many
individual features make Equalizer a unique, open source framework to develop
visualization applications for virtually any type of setup and use case.
While individual applications and case studies using the framework, as well as
new algorithms and system components extending it have been presented
since the initial release of Equalizer, many important features and functionalities
have not previously been published and are presented here.
In this report, we thus present an updated comprehensive review of the
integration of research, application use cases and commercial developments
with respect to Equalizer as well as novel comparative experimental results
of its scalability features.

We present novel algorithms for parallel rendering which did not appear in a
separate publication, including pixel and sub-pixel decompositions, dynamic
frame resolution, tunable sort-first load-balancing parameters, frame-rate
equalization, thread synchronization modes for multi-GPU rendering, a powerful
abstraction for multi-view rendering on arbitrary display setups, dynamic focus
distance and asymmetric eye positions for VR, parallel pixel streaming to tiled
display walls, as well as a fully-fledged data distribution API with compression
and reliable multicast.

The remainder of this paper is structured as follows: First we provide an update
on related work since the introduction of \textsf{Equalizer}. The main body of this paper
then presents new performance features, VR algorithms, usability
features to build complex applications, main novel features of the underlying
\textsf{Collage} network library and a quick overview of the main
\textsf{Equalizer}-based applications. A result section presents new experiments
not previously published, followed by the discussion and conclusion.

\section{Related Work}\label{sec:related}

In 2009 we presented \textsf{Equalizer}~\cite{EMP:09}, which introduced the
architecture of a generic parallel rendering framework and summarized our work
in parallel rendering. Since then, an extensive Programming and User Guide
provides in-depth documentation on using and programming
\textsf{Equalizer}~\cite{Eilemann:13}. In the following related work we assume
these two publications and their references as a starting point, and focus on
the new work published since 2009.

The concept of transparent OpenGL interception popularized by \textsf{WireGL}
and \textsf{Chromium}~\cite{HHNFAKK:02} has received little attention since
2009. While some commercial implementations such as \textsf{TechViz} and
\textsf{MechDyne Conduit} continue to exist, on the research side only
\textsf{ClusterGL}~\cite{NHM:11} has been presented. \textsf{ClusterGL} employs
the same approach as \textsf{Chromium}, but delivers a significantly faster
implementation of transparent OpenGL interception and distribution for parallel
rendering. \textsf{CGLX}~\cite{DK:11} tries to bring parallel execution
transparently to OpenGL applications, by emulating the GLUT API and intercepting
certain OpenGL calls. In contrast to frameworks like \textsf{Chromium} and
\textsf{ClusterGL} which distribute OpenGL calls, \textsf{CGLX} follows the
distributed application approach. This works transparently for trivial
applications, but quickly requires the application developer to address the
complexities of a distributed application, when mutable application state needs
to be synchronized across processes. For realistic applications, writing
parallel applications remains the only viable approach for scalable parallel
rendering, as shown by the success of \textsf{Paraview}, \textsf{Visit} and
various \textsf{Equalizer}-based applications.

On the other hand, software for driving and interacting with tiled display walls
has received significant attention, including \textsf{Sage}~\cite{Sage} and
\textsf{Sage~2}~\cite{Sage2} in particular. \textsf{Sage} was built entirely
around the concept of a shared framebuffer where all content windows are
separate applications using pixel streaming. It is no longer actively supported.
\textsf{Sage~2} is a complete, browser-centric reimplementation where each
application is a web application distributed across browser instances.
\textsf{DisplayCluster}~\cite{DC}, and its continuation
\textsf{Tide}~\cite{tide}, also implement the shared framebuffer concept of
\textsf{Sage}, but provide a few native content applications integrated into the
display servers. All these solutions implement a scalable display environment and
are a target display platform for scalable 3D graphics applications.

\textsf{Equalizer} itself has received significant attention within the research
community. Various algorithms to improve the parallel rendering performance have
been proposed: compression and region of interest during
compositing~\cite{MEP:10}, load-balancing resources for multi-display
installations~\cite{EEP:11}, asynchronous compositing and NUMA
optimizations~\cite{EBAHMP:12}, as well as work queueing~\cite{SPEP:16}.
Additionally, complex large scale and out-of-core multiresolution
rendering approaches have been parallelized and implemented with
Equalizer~\cite{GMBP:10, GEMPG:13}, demonstrating the feasibility of
the framework to be used with complex rendering algorithms and 3D model
representations.

Furthermore, various applications and frameworks have used \textsf{Equalizer}
for new research in visualization. On the application side, \textsf{RTT
Deltagen}, \textsf{Bino}, \textsf{Livre} and \textsf{RTNeuron}~\cite{HBBES:13}
are the most mature examples and are presented in \sref{sApplications}. On the
framework side, \textsf{Omegalib}~\cite{Omegalib}, a framework used in the
Cave2, made significant progress in integrating 2D collaborative workspaces like
\textsf{Sage~2} with 3D immersive content. Lambers et.al. developed a framework
for visualizing remote sensing data~\cite{LK:09} on large displays and immersive
installations.

\section{Performance Features}


\subsection{New Decomposition Modes}

\begin{figure*}[ht]\center
  \begin{subfigure}[b]{0.24\textwidth}
    \includegraphics[width=\textwidth]{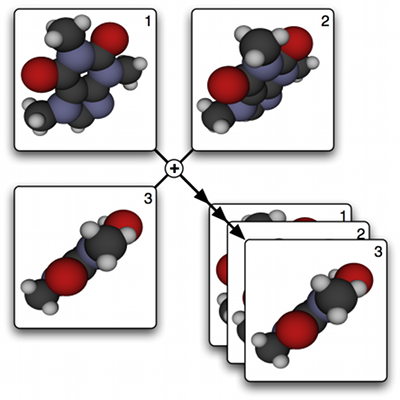}
  \end{subfigure}
  \begin{subfigure}[b]{0.24\textwidth}
    \includegraphics[width=\textwidth]{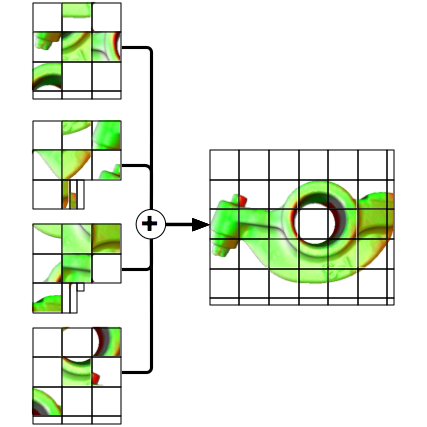}
  \end{subfigure}
  \begin{subfigure}[b]{0.24\textwidth}
    \includegraphics[width=\textwidth]{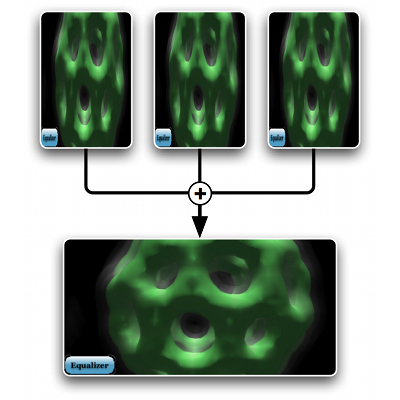}
  \end{subfigure}
  \begin{subfigure}[b]{0.24\textwidth}
    \includegraphics[width=\textwidth]{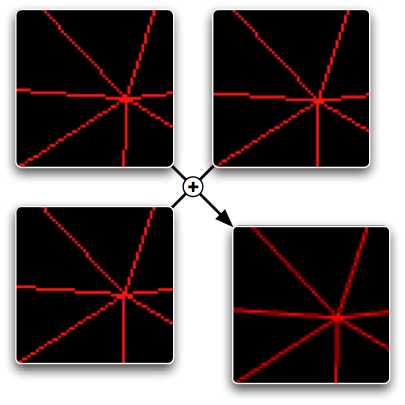}
  \end{subfigure}\\[\medskipamount]
  \begin{subfigure}[t]{0.24\textwidth}
    {\tiny\begin{lstlisting}
compound {
  channel "destination"
  framerate_equalizer {}
  compound {
    channel "source1"
    phase 0 period 3
    outputframe  { name "frame" }
  }
  compound {
    channel "source2"
    phase 1 period 3
    outputframe  { name "frame" }
  }
  compound {
    channel "source3"
    phase 2 period 3
    outputframe  { name "frame" }
  }
  inputframe { name "frame" }
}
    \end{lstlisting}\vspace{24ex}}
    \caption{\label{fig:dplex}Time-Multiplex}
  \end{subfigure}
  \begin{subfigure}[t]{0.24\textwidth}
    {\tiny\begin{lstlisting}
compound {
  channel "destination"
  outputtiles {
    name "queue"
    size [ 64 64 ]
  }
  compound {
    channel "destination"
    inputtiles { name "queue" }
  }
  compound {
    channel "source1"
    inputtiles { name "queue" }
    outputframe {}
  }
  compound {
    channel "source2"
    inputtiles { name "queue" }
    outputframe {}
  }
  compound {
    channel "source3"
    inputtiles { name "queue" }
    outputframe {}
  }
  inputframe { name "frame.source1" }
  inputframe { name "frame.source2" }
  inputframe { name "frame.source3" }
}
    \end{lstlisting}}
    \caption{\label{fig:tiles}Tiles}
  \end{subfigure}
  \begin{subfigure}[t]{0.24\textwidth}
    {\tiny\begin{lstlisting}
compound {
  channel "dest"
  compound {
    channel "dest"
    pixel [ 0 0 3 1 ]
    outputframe { type texture }
  }
  compound {
    channel "source1"
    pixel [ 1 0 3 1 ]
    outputframe {}
  }
  compound {
    channel "source2"
    pixel [ 2 0 3 1 ]
    outputframe {}
  }
  inputframe { name "frame.dest" }
  inputframe { name "frame.source1" }
  inputframe { name "frame.source2" }
}
    \end{lstlisting}\vspace{21ex}}
    \caption{\label{fig:pixel}Pixel}
  \end{subfigure}
  \begin{subfigure}[t]{0.24\textwidth}
    {\tiny\begin{lstlisting}
compound {
  channel "dest"
  compound {
    channel "dest"
    subpixel [ 0 3 ]
    outputframe { type texture }
  }
  compound {
    channel "source1"
    subpixel [ 1 3 ]
    outputframe {}
  }
  compound {
    channel "source2"
    subpixel [ 1 3 ]
    outputframe {}
  }
  inputframe { name "frame.dest" }
  inputframe { name "frame.source1" }
  inputframe { name "frame.source2" }
}
    \end{lstlisting}\vspace{21ex}}
    \caption{\label{fig:subpixel}Subpixel}
  \end{subfigure}
  \caption{New \textsf{Equalizer} task decomposition modes and their compound
descriptions for parallel rendering}
  \label{fig:compounds}
\end{figure*}

The initial version of \textsf{Equalizer} implemented the basic sort-first (2D),
sort-last (DB), stereo (EYE) and multilevel decompositions \cite{EMP:09}. In the
following we present the newly added decomposition modes and motivate their use
case, which bring the overall feature set way beyond the typical sort-first and
sort-last rendering modes. \fig{fig:compounds} provides an overview of the new
modes. The compound concept to set up scalable rendering is presented in
\cite{EMP:09}.

\subsubsection{Time-Multiplex}

Time-multiplexing (\fig{fig:dplex}), also called AFR or DPlex, was first
implemented in \cite{BRE:05} for shared memory machines. It is however a better
fit for distributed memory systems, since the separate memory space makes
concurrent rendering of different frames easier to implement. While it increases
the framerate linearly, it does not decrease the latency between user input and
the corresponding output. Consequently, this decomposition mode is mostly useful
for non-interactive movie generation. It is transparent to \textsf{Equalizer}
applications, but does require the configuration latency to be equal or greater
than the number of source channels. Furthermore, to work in multi-threaded,
multi-GPU configurations, the application needs to support running the rendering
threads asynchronously (\sref{sec:threading}). The output frame rate of the
destination channel may be smoothened using a {\em frame rate equalizer}
(\sref{sec:framerateEq}).

\subsubsection{Tiles and Chunks}\label{sec:tile}

Tile (\fig{fig:tiles}) and chunk decompositions are a variant of sort-first and
sort-last rendering, respectively. They decompose the scene into a predefined
set of fixed-size image tiles or database ranges. These tasks, or work packages,
are queued and processed by all source channels by polling a server-central
queue. Prefetching ensures that the task communication overlaps with rendering.
As shown in~\cite{SPEP:16} and the results, these modes can provide better
performance due to being implicitly, i.e., inherently load-balanced, as long as
there is an insignificant overhead for the render task setup. This mode is
transparent to \textsf{Equalizer} applications.

\subsubsection{Pixel}

Pixel compounds (\fig{fig:pixel}) decompose the destination channel by
interleaving rows or columns in image space. They are a variant of sort-first
decomposition which works well for fill-limited applications which are not
geometry bound, for example direct volume rendering. Source channels cannot
reduce geometry load through view frustum culling, since each source channel has
almost the same frustum. However, the fragment load on all source channels is
reduced linearly and well load-balanced due to the interleaved distribution of
pixels. This functionality is transparent to \textsf{Equalizer} applications,
and the default compositing uses the stencil buffer to blit pixels onto the
destination channel.

\subsubsection{Subpixel}

Subpixel compounds (\fig{fig:subpixel}) are similar to pixel compounds, but they
decompose the work for a single pixel, for example with Monte-Carlo ray tracing,
FSAA or depth of field rendering. Composition typically uses accumulation and
averaging of all computed fragments for a pixel. This feature is not fully
transparent to the application, since it needs to adapt (jitter or tilt) the
frustum based on the iteration executed.


\subsection{Equalizers}

{\em Equalizers} are an addition to compound trees. They modify parameters of
their respective subtree at runtime to dynamically optimize the resource usage,
by each tuning one aspect of the decomposition. Due to their nature, they are
transparent to application developers, but might have application-accessible
parameters to tune their behavior. Resource equalization is the critical
component for scalable parallel rendering, and therefore the eponym for the
\textsf{Equalizer} project name.

\begin{figure*}[ht]\center
  \begin{subfigure}[b]{0.24\textwidth}
    \includegraphics[width=\textwidth]{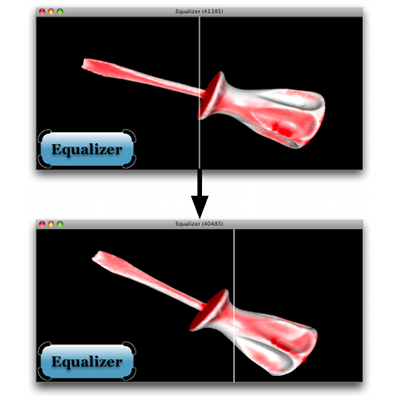}
      \caption{\label{fig:loadeq}Load-Balancing}
  \end{subfigure}
  \begin{subfigure}[b]{0.24\textwidth}
    \includegraphics[width=\textwidth]{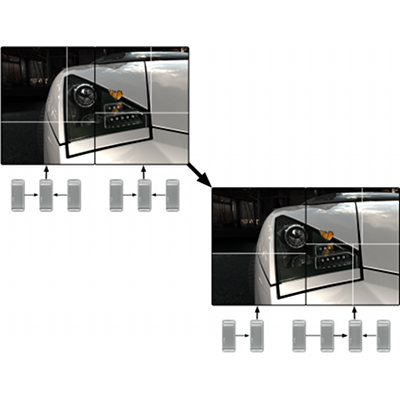}
      \caption{\label{fig:vieweq}Cross-Segment Load-Balancing}
  \end{subfigure}
  \begin{subfigure}[b]{0.24\textwidth}
    \includegraphics[width=\textwidth]{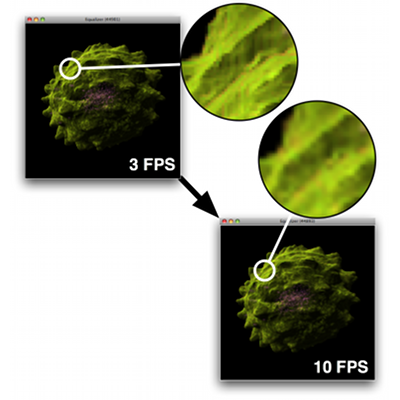}
      \caption{\label{fig:dfr}Dynamic Frame Resolution}
  \end{subfigure}
  \begin{subfigure}[b]{0.24\textwidth}
    \includegraphics[width=\textwidth]{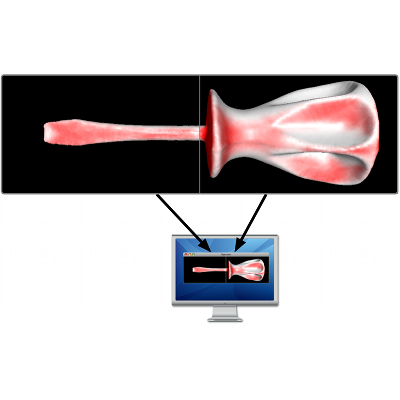}
      \caption{\label{fig:monitor}Monitoring}
  \end{subfigure}
  \caption{Runtime modifications}
  \label{fig:equalizers}
\end{figure*}

\subsubsection{Sort-First and Sort-Last Load Balancing}

Sort-first (\fig{fig:loadeq}) and sort-last load balancing are the most obvious
optimizations for these parallel rendering modes. Our load equalizers are fully
transparent for application developers; that is, they use a reactive approach
based on past rendering times. This assumes a reasonable frame-to-frame
coherence. Equalizer implements two different algorithms, a
\textsf{load\_equalizer} and a \textsf{tree\_equalizer}. The result section
provides some evidence on the strengths and weaknesses of both algorithms.

The \textsf{load\_equalizer} stores a 2D or 1D grid of the load, mapping the
load of each channel. The load is stored in normalized 2D/1D coordinates using
$\frac{time}{area}$ as the load, the contributing source channels are organized
in a binary tree, and then the algorithm balances the two branches of each level
by equalizing the integral over the cost area map on each side.

The \textsf{tree\_equalizer} also uses a binary tree for recursive load
balancing. It computes the accumulated render time on all nodes of the tree, and
uses this to allocate an equal render time to each subtree. It makes no
assumption of the load distribution in 2D or 1D space, it only tries to correct
the imbalance in render time.

Both equalizers implement tuneable parameters allowing application developers to
optimize the load balancing based on the characteristics of their rendering
algorithm:

\begin{compactdesc}
\item[Split Mode] configures the tile layout: horizontal or vertical stripes,
and 2D, a binary tree split alternating the split axis on each level, resulting
in compact 2D tiles.
\item[Damping] reduces frame-to-frame oscillations. The equal load distribution
within the region of interest assumed by the load equalizer is in reality not
equal, causing the load balancing to overshoot. Damping is a normalized scalar
defining how much of the computed delta from the previous position is applied to
the new split.
\item[Resistance] eliminates small deltas in the load balancing step. This might
help the application to cache visibility computations since the frustum does not
change each frame.
\item[Boundaries] define the modulo factor in pixels onto which a load split may
fall. Some rendering algorithms produce artefacts related to the OpenGL raster
position, e.g., screen door transparency, which can be eliminated by aligning
the boundary to the pixel repetition. Furthermore, some rendering algorithms are
sensitive to cache alignments, which can again be exploited by chosing the
corresponding boundary.
\end{compactdesc}

\subsubsection{Dynamic Work Packages}

The package equalizers implement client-affinity models for tile and chunk
compounds (\sref{sec:tile}). A \textsf{tile\_equalizer} or
\textsf{chunk\_equalizer} creates the packages and changes the assignment of them
to individual nodes, based on an affinity model specified in the equalizer.
In~\cite{SPEP:16}, we explore this approach in detail.

\subsubsection{Cross-Segment Load Balancing}

Cross-segment load balancing (\fig{fig:vieweq}) addresses the optimal resource
allocation of $n$ rendering resources to $m$ output channels (with $n\geq m$). A
view equalizer works in conjunction with load equalizers balancing the
individual output channels. It monitors the usage of shared source channels
across outputs and activates them to balance the rendering time of all outputs.
In \cite{EEP:11}, we provide a detailed description and evaluation of our
algorithm.

\subsubsection{Dynamic Frame Resolution}

The DFR equalizer (\fig{fig:dfr}) provides a functionality similar to dynamic
video resizing \cite{MBDM:97}, that is, it maintains a constant framerate by
adapting the rendering resolution of a fill-limited application. In
\textsf{Equalizer}, this works by rendering into a source channel (typically on
a FBO) separate to the destination channel, and then scaling the rendering
during the transfer (typically through an on-GPU texture) to the destination
channel. The DFR equalizer monitors the rendering performance and accordingly
adapts the resolution of the source channel and zoom factor for the source to
destination transfer. If the performance and source channel resolutions allow,
this will not only subsample, but also supersample the destination channel to
reduce aliasing artefacts.

\subsubsection{Frame Rate Equalizer}\label{sec:framerateEq}

The framerate equalizer smoothens the output frame rate of a destination channel
by instructing the corresponding window to delay its buffer swap to a minimum
time between swaps. This is regularly used for time-multiplexed decompositions,
where source channels tend to drift and finish their rendering unevenly
distributed over time. This equalizer is however fully independent of DPlex
compounds, and may be used to smoothen irregular application rendering
algorithms.

\subsubsection{Monitoring}

The monitor equalizer (\fig{fig:monitor}, \fig{fLayout}) allows to reuse the
rendering on another channel, typically for monitoring a larger setup on a
control workstation. Output frames on the display channels are connected to
input frames on a single monitoring channel. The monitor equalizer changes the
scaling factor and offset between the output and input, so that the monitor
channel has the same, but typically downscaled view, as the originating
segments.

\subsection{Optimizations}

\subsubsection{Region of Interest}

The region of interest is the screen-space 2D bounding box enclosing the data
rendered by a single resource. We have extended the core parallel rendering
framework to use an application-provided ROI to optimize the
\textsf{load\_equalizer} as well as the image compositing. The load equalizer
uses the ROI to refine its load grid to the regions containing data. The
compositing code uses the ROI to minimize image readback and network
transmission. In \cite{MEP:10} and \cite{EBAHMP:12}, we provide the details of
the algorithm, and show that using ROI can quadruple the rendering performance,
in particular for the costly compositing step in sort-last rendering.

\subsubsection{Asynchronous Compositing}

Asynchronous compositing is pipelining the rendering with compositing operations, by
executing the image readback, network transfer and image assembly from threads
running in parallel to the rendering threads. In \cite{EBAHMP:12}, we provide
the details of the implementation and experimental data showing an improvement
of the rendering performance of over 25\% for large node counts.

\subsubsection{Download and Compression Plugins}

Compression for the compositing step is critical for performance. This not only
applies to the well-researched network transfer step, but also for the transfer
between GPU and CPU. \textsf{Equalizer} supports a variety of compression
algorithms, from very fast run-length encoding (RLE) and YUV subsampling on the GPU to JPEG
compression. These algorithms are implemented as runtime-loaded plugins,
allowing easy extension and customization to application-specific compression.
In \cite{MEP:10}, we show this to be a critical step for interactive performance
at scale.

\subsubsection{Thread Synchronization Modes}\label{sec:threading}

Different applications have different degrees on how decoupled and thread-safe
the rendering code is from the application logic. For full decoupling all
mutable data has to have a copy in each render thread, which is not feasible in
most applications and large data scenarios. To easily customize the
synchronization of all threads on a single process, \textsf{Equalizer}
implements three threading modes:  Full synchronization, draw synchronization
and asynchronous. Note that the execution between node processes is always
asynchronous, for up to \textsf{latency} frames.

In full synchronization, all threads always execute the same frame; that is, the
render threads are unlocked after \textsf{Node::frameStart}, and the node is
blocked for all render threads to finish the frame before executing
\textsf{Node::frameFinish}. This allows the render threads to read shared data
from all their operations, but provides the slowest performance.

In draw synchronization, the node thread and all render threads are synchronized
for all \textsf{frameDraw} operations; that is, \textsf{Node::frameFinish} is
executed after the last channel is done drawing. This allows the render threads
to read shared data during their draw operation, but not during compositing.
Since compositing is often independent of the rendered data, this is the default
mode. This mode allows to overlap compositing with rendering and data
synchronization on multi-GPU machines.

In asynchronous execution, all threads run asynchronously. Render threads may
work on different frames at any given time. This mode is the fastest, and
requires the application to have one instance of each mutable object in each
render thread. It is required for scaling time-multiplex compounds on multi-GPU
machines.

\section{Virtual Reality Features}

Virtual Reality is an important field for parallel rendering. It does however
require special attention to support it as a first-class citizen in a generic
parallel rendering framework. \textsf{Equalizer} has been used in many virtual
reality installations, such as the Cave2 \cite{FNTTL:13}, the high-resolution C6
CAVE at the KAUST visualization laboratory, and head-mounted displays
(\fig{FIG_teaser}). In the following we lay out the features needed to support
these installations, motivated by application use cases.

\subsection{Head Tracking}

Head tracking is the minimal feature needed to support immersive installations.
\textsf{Equalizer} does support multiple, independent tracked views through the
observer abstraction (\sref{sec:observer}). Built-in VRPN support enables the
direct, application-transparent configuration of a VRPN tracker device.
Alternatively, applications can provide a $4\times 4$ tracking matrix. Both
CAVE-like tracking with fixed projection surfaces and HMD tracking with moving
displays are implemented.

\subsection{Dynamic Focus Distance}

To our knowledge, all parallel rendering systems have the focal plane coincide
with the physical display surface. For better viewing comfort, we introduce a
new dynamic focus mode, where the application defines the distance of the focal
plane from the observer, based on the current \textit{lookat} distance. Initial
experiments show that this provides better viewing comfort, in particular for
objects placed in front of the physical displays.

\subsection{Asymmetric Eye Position}

Traditional head tracking computes the left and right eye positions by using an
interocular distance. However, human heads are not symmetric, and by measuring
individual users a more precise frustum can be computed. \textsf{Equalizer}
supports this through the optional configuration of individual 3D eye
translations relative to the tracking matrix.

\subsection{Model Unit}

This feature allows applications to specify a scaling factor between the model
and the real world, to allow exploration of macroscopic or microscopic worlds in
virtual reality. The unit is per view, allowing different scale factors within
the same application. It scales both the specified projection surface as well
as the eye position (and therefore separation) to achieve the necessary effect.

\subsection{Runtime Stereo Switch}

Applications can switch each view between mono and stereo rendering at runtime,
and run both monoscopic and stereoscopic views concurrently (\fig{FIG_teaser}
(b)). This switch does potentially involve the start and stop of resources and
processes for passive stereo or stereo-dependent task decompositions
(\sref{sec:reconfig}).

\section{Usability Features}

In this section we present features motivated by real-world application use
cases, i.e., new functionalities rather then performance improvements. We
motivate the use case, explain the architecture and integration into our
parallel rendering framework, and, where applicable, show the steps needed to
use this functionality in applications.

\subsection{Physical and Logical Visualization Setup}

Real-world visualization setups can be complex. An abstract representation of
the display system simplifies the configuration process. Applications often have
the need to be aware of spatial relationships of the display setup, for example
to render 2D overlays or to configure multiple views on a tiled display wall.

We addressed this need through a new configuration section interspersed between
the node/pipe/window/channel hardware resources and the compound trees
configurating the resource usage for parallel rendering.

A typical installation consists of one projection canvas, which is one
aggregated projection surface, e.g., a tiled display wall or a CAVE. Desktop
windows are considered a canvas. Each canvas is made of one or more segments,
which are the individual outputs connected to a display or projector. Segments
can be planar or non-planar to each other, and can overlap or have gaps between
each other. A segment is referencing a channel, which defines the output area of
this segment, e.g., on a DVI connector connected to a projector. This
abstraction covers all use cases from simple windows, tiled display walls with
bezels, to non-planar immersive systems with edge-blending.

A canvas can define a frustum, which will create default, planar sub-frusta for
all of its segments. A segment can also define a frustum, which overrides the
canvas frustum, e.g., for non-planar setups such as CAVEs or curved screens.
These frusta describe a physically-correct display setup for a Virtual Reality
installation. \label{sec:swap} A canvas may have a software or hardware swap
barrier, which will synchronize the rendering of all contributing GPUs. The
software barrier executes a \textsf{glFinish} to ensure the GPU is ready to
swap, a \textsf{Collage} barrier (\sref{sec:barrier}) to synchronize all
segments, and the swap buffers call followed by a \textsf{glFlush} to ensure timely
execution of the swap command. The hardware swap barrier is implemented using
the \textsf{NV\_swap\_group} extension.

On each canvas, the application can display one or more views. A view is used in
the sense of the MVC pattern. The view class is available to applications to
define view-specific data for rendering, e.g., a scene, viewing mode or camera.
The application process manages this data, and the render clients receive it for
rendering.

A layout groups one or more views which logically belong together. A layout is
applied to a canvas. The layout assignment can be changed at run-time by the
application. The intersection between views and segments defines which output
channels are available, and which frustum they should use for rendering. These
output channels are then used as destination channels in a compound. They are
automatically created during configuration.

\label{sec:observer}
An observer looks at one or more views. It is described by the observer position
in the world and its eye separation. Each observer has its own stereo
mode, focus distance and eye positions. This allows to have untracked
views and multiple tracked views, e.g., two HMDs, in the same application.

\begin{figure}[ht]\center
  \includegraphics[width=\columnwidth]{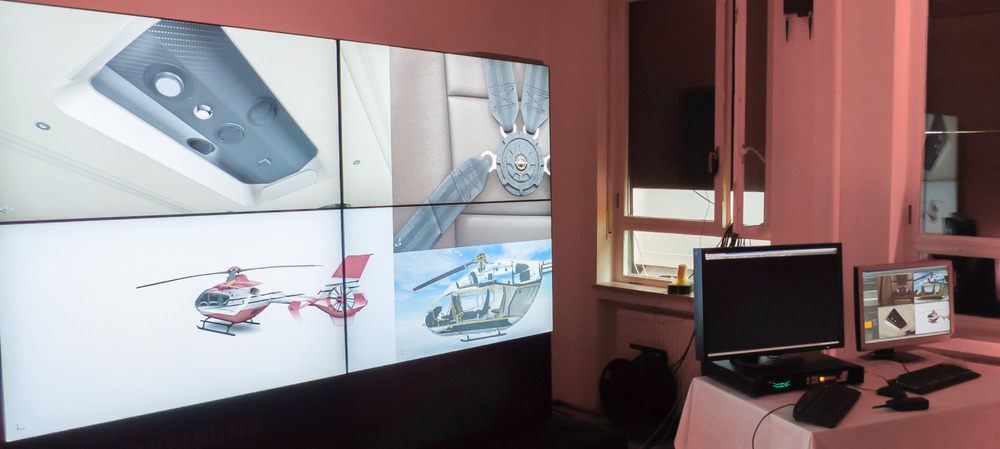}
  \caption{\label{fLayout}A 2x2 tiled display wall and control host rendering four independent views driven by an eight node visualization cluster}
\end{figure}

\fig{fLayout} shows RTT Deltagen running an example multi-segment, multi-view
setup driven by eight rendering nodes. The main tiled display wall canvas uses
four LCD segments showing one layout with four views, which do not align on the
segment boundaries. This setup creates seven destination channels. The
configuration provides multiple, run-time configurable layouts. It is driven
from the control host on the right, which shows four views, each in their own
canvas and segment windows with a single-view layout each. One view on the
control host synchronizes its content (model and camera) with one view on the
display wall through Collage objects. The control host allows full model
modifications and all workflows supported within the standalone Deltagen
application, and all changes are synchronized to the corresponding rendering
nodes. For this monoscopic setup no head tracking or observers are used.

\subsection{Runtime Reconfiguration}\label{sec:reconfig}

Switching a layout, as described above, or switching the stereo rendering mode,
may involve a different set of resources after the change, including the launch
and exit of render client processes. \textsf{Equalizer} solves this through a
reconfiguration step at the beginning of each rendering frame. Each resource
(channel, window, pipe, node) has an activation count, which is updated when the
layout or any other relevant rendering parameter is changed. When a resource is
found whose activation count does not match its current start/stopped state, the
resource is created or destroyed and \textsf{configInit} or \textsf{configExit}
are called accordingly. In the current implementation, a normal configuration
initialization or exit, as described in \cite{EMP:09}, uses the same code path
with all used resources transitioning to a running or stopped state,
accordingly. Since starting new resources typically requires object mapping and
associated data distribution, it is a costly operation.

\subsection{Automatic Configuration}

Automatic configuration implements the discovery of local and remote resources
as well as the creation of typical configurations using the discovered resources
at application launch time.

The discovery is implemented in a separate library, \textsf{hwsd} (HardWare
Service
Discovery), which uses a plugin-based approach to discover GPUs for GLX, AGL or
WGL windowing systems, as well as network interfaces on Linux, Mac OS X and
Windows. Furthermore, it detects the presence of VirtualGL to allow optimal
configuration of remote visualization clusters. The resources can be discovered
on the local workstation, and through the help of a simple daemon using the
zeroconf protocol, on a set of remote nodes within a visualization cluster. A
session identifier may be used to support multiple users on a single cluster.

The \textsf{Equalizer} server uses the hwsd library to discover local and remote
resources when an hwsd session name instead of a \texttt{.eqc} configuration
file is provided. A set of standard decomposition modes is configured, which can
be selected through activating the corresponding layout.

This versatile mechanism allows non-experts to configure and profit from
multi-GPU workstations and visualization clusters, as well as to provide system
administrators with the tools to implement easy to use integration with cluster
schedulers.

\subsection{Qt Windowing}

Qt is a popular window system with application developers. Unfortunately, it
imposes a different threading model for window creation and event handling
compared to \textsf{Equalizer}. In \textsf{Equalizer}, each GPU rendering thread
is independently responsible for creating its windows, receiving the events and
eventually dispatching them to the application process' main thread. This design
is motivated by the natural threading model of X11 and WGL, and allows simple
sequential semantics between OpenGL rendering and event handling. In contrast,
Qt requires all windows and each QOpenGLContext to be created from the Qt main
thread. An existing Qt window or context may subsequently be moved to a
different thread, and events are signalled from the main thread. For Qt windows,
Equalizer will internally dispatch and handle the window creation from the
render to the main thread, move the created objects back to the render thread,
and dispatch Qt signals to the correct render threads.

\subsection{Tide Integration}

\textsf{Tide} (Tiled interactive display environment) is an improved version of
\textsf{DisplayCluster}~\cite{DC}, providing a touch-based, multi-window user
interface for high-resolution tiled display walls. Remote applications receive
input events and send pixel streams using the \textsf{Deflect} client library.
\textsf{Equalizer} includes full support, enabling application-transparent
integration with \textsf{Tide}. When a \textsf{Tide} server is configured, all
output channels of a view stream in parallel to one window on the wall. In
\cite{deflect}, we have shown interactive framerates for a 24 megapixel
resolution over a WAN link. \textsf{Deflect} events are translated and injected
into the \textsf{Equalizer} event flow, allowing seamless application
integration.

\subsection{\textsf{Sequel}}\label{sec:sequel}

\textsf{Sequel} is a simplification layer for \textsf{Equalizer}. It is based on
the realization that while fully expressive, the verbatim abstraction layer of
nodes, pipes, windows and channels in \textsf{Equalizer} requires significant
learning to fully understand and exploit. In reality, a higher abstraction of
only \textsf{Application} and \textsf{Renderer} is sufficient for many use
cases. In \textsf{Sequel}, the application class drives the configuration, and
one renderer instance is created for each (pipe) render thread. They also
provide the natural place to store and distribute data. Finally,
\textsf{ViewData} provides a convenient way to manage multiple views by storing
the camera, model or any other view-specific information.

\section{The \textsf{Collage} Network Library}

An important part of writing a parallel rendering application is the
communication layer between the individual processes. \textsf{Equalizer} relies
on the
\textsf{Collage} network library for its internal operation. \textsf{Collage}
furthermore provides
powerful abstractions for writing \textsf{Equalizer} applications, which are
introduced
in this section.

\subsection{Architecture}

\begin{figure}[ht]\center
  \includegraphics[width=\columnwidth]{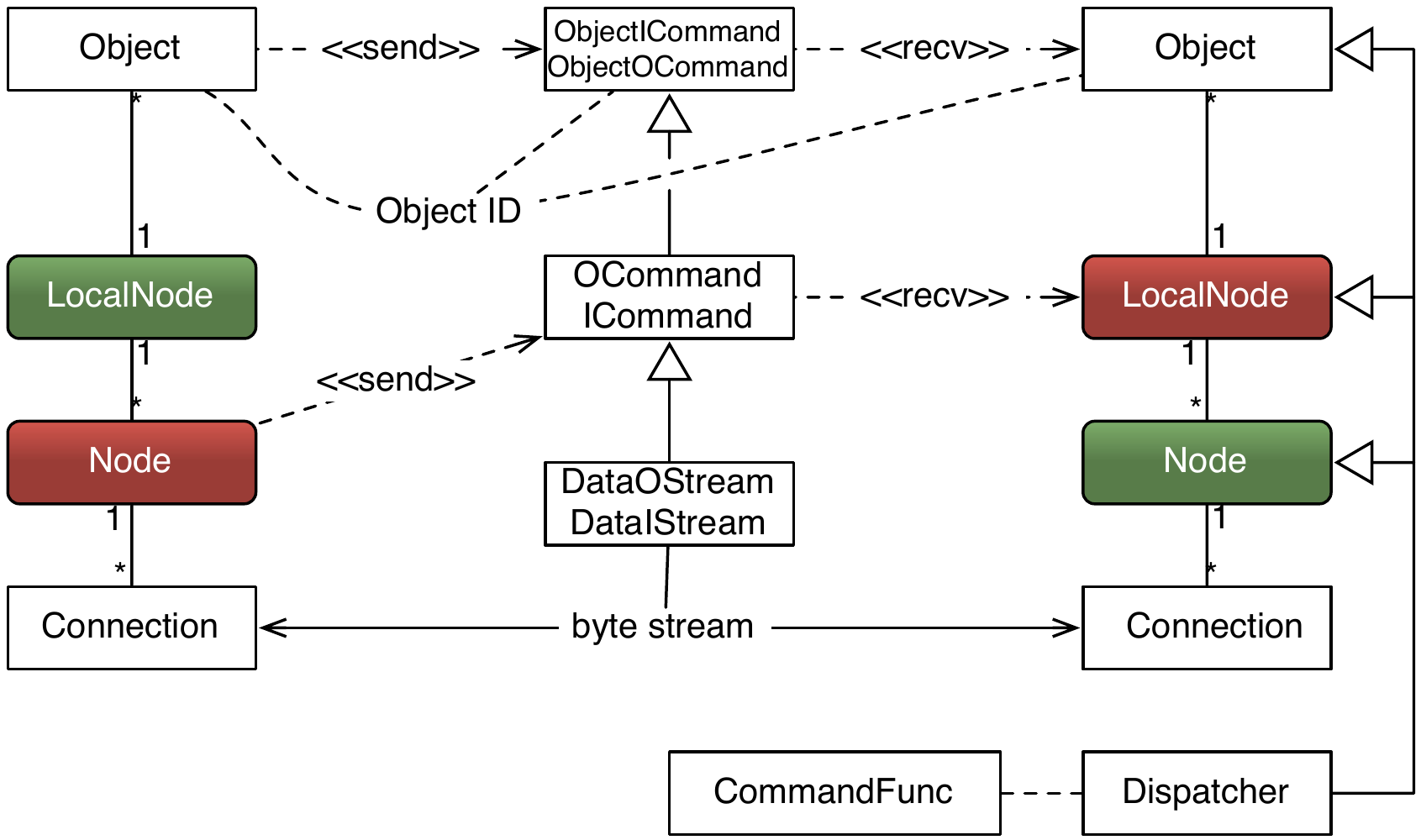}
  \caption{\label{fNetObject}Communication between two \textsf{Collage} objects}
\end{figure}

\textsf{Collage} provides networking functionality of different abstraction
layers, gradually providing higher level functionality for the programmer. The
main primitives in \textsf{Collage} and their relations are shown in
\fig{fNetObject} and provide:

\begin{compactdesc}
\item[Connection] A stream-oriented point-to-point communication
  line. The connections
  transmit raw data reliably between two endpoints for unicast connections, and
  between a set of endpoints for multicast connections. For unicast,
  process-local pipes, TCP and Infiniband RDMA are implemented. For multicast,
  a reliable, UDP-based protocol is discussed in \sref{sec:RSP}.
\item[DataI/OStream] Abstracts the input and output of C++ data types from or to
  a set of connections by implementing output stream operators. Uses buffering
  to aggregate data for network transmission. Performs byte swapping during
  input if the endianness differs between the remote and local node.
\item[Node and LocalNode] The abstraction of a process in the cluster. Nodes
  communicate with each other using connections. A LocalNode listens on various
  connections and processes requests for a given process. Received data is
  wrapped in ICommands and dispatched to command handler methods. A Node is a
  proxy for a remote LocalNode. The \textsf{Equalizer} Client object is a
LocalNode.
\item[Object] Provides object-oriented, versioned data distribution of C++
  objects between nodes. Objects are registered or mapped on a Local\-Node.
\end{compactdesc}

\subsection{Reliable Stream Protocol}\label{sec:RSP}

RSP is an implementation of a reliable multicast protocol over unreliable UDP
transport. RSP behaves similarly to TCP; it provides full reliability and
ordering of the data, and slow receivers will eventually throttle the sender
through a sliding window algorithm. This behavior is needed to guarantee
delivery of data in all situations. Pragmatic generic multicast (PGM~\cite{pgm})
provides full ordering, but slow clients will disconnect from the multicast
session instead of throttling the send rate.

RSP combines various established algorithms~\cite{adamson2004negative,
  Gau:2002:MFC:506824.506832} for multicast in an open source implementation
capable of delivering wire speed transmission rates on high-speed LAN
interfaces. In the following we will outline the RSP protocol and implementation
as well as motivate the design decisions. Any defaults given below are for
Linux or OS X, the Windows UDP stack requires different default values which can
be found in the implementation.

Our RSP implementation uses a separate protocol thread for each RSP group, which
handles all reads and writes on the multicast socket. It implements the protocol
handling and communicates with the application threads through thread-safe
queues. The queues contain datagrams of the application byte stream, prefixed by
a header of at most eight bytes. Each connection has a configurable number of
buffers (1024 by default) of a configurable MTU (1470 bytes default), which are
either free or in transmission.

Handling a smooth packet flow is critical for performance. RSP uses active flow
control to advance the byte stream buffered by the implementation. Each incoming
connection actively acknowledges every $n$ (17 by default) packets fully
received. The incoming connections offset this acknowledgment by their
connection identifier to avoid bursts of acks. Any missed datagram is actively
nack'ed as soon as detected. Write connections continuously retransmit packets
for nack datagrams, and advance their window upon reception of all acks from the
group. The writer will explicitly request an ack or nack when it runs out of
empty buffers or finishes its write queue. Nack datagrams may contain multiple
ranges of missed datagrams, which is motivated by the observation that UDP
implementations often drop multiple contiguous packets.

Congestion control is necessary to optimize bandwidth usage. While TCP uses the
well-known additive increase, multiplicative decrease algorithm, we have chosen
a more aggressive congestion control algorithm of additive increase and additive
decrease. This has proven experimentally to be more optimal: UDP is often
rate-limited by switches; that is, packets are discarded regularly and not
exceptionally. Only slowly backing of the current send rate helps to stay close
to this limit. Furthermore, our RSP traffic is limited to the local subnet,
making cooperation between multiple data stream less of an issue. Send rate
limiting uses a bucket algorithm, where over time the bucket fills with send
credits, from which sends are substracted. If there are no available credits,
the sender sleeps until sufficient credits are available.

\subsection{Distributed, Versioned Objects}

Adapting an existing application for parallel rendering requires the
synchronization of application data across the processes in the parallel
rendering setup. Existing parallel rendering frameworks address this often
poorly, at best they rely on MPI to distribute data. Real-world, interactive
visualization applications are typically written in C++ and have complex data
models and class hierarchies to represent their application state. As outlined
in \cite{EMP:09}, the parallel rendering code in an \textsf{Equalizer}
application only needs access to the data needed for rendering, as all
application logic is centralized in the application main thread. We have
encountered two main approaches to address this distribution: Using a shared
filesystem for static data, or using data distribution for static and dynamic
data. Distributed objects are not required to build \textsf{Equalizer}
applications. While most developers choose to use this abstraction for
convenience, we have seen applications using other means for data distribution,
e.g., MPI.

\subsubsection{Programming Interface}

Distributed objects in \textsf{Collage} provide powerful, object-oriented data
distribution for C++ objects. They facilitate the implementation of data
distribution in a cluster environment. Distributed objects are created by
subclassing from \textsf{co::Serializable} or \textsf{co::Object}. The
application programmer implements serialization and deserialization. Distributed
objects can be static (immutable) or dynamic. Objects have a universally unique
identifier (UUID) as cluster-wide address. A master-slave model is used to
establish mapping and data synchronization across processes. Typically, the
application main loop registers a master instance and communicates the UUID to
the render clients, which map their instance to the given identifier. The
following object types are available:

\begin{compactdesc}
\item[Static] The object is not versioned nor buffered. The instance data is
  serialized whenever a new slave instance is mapped. No additional data is
  stored.
\item[Instance] The object is versioned and buffered. The instance and delta
  data are identical; that is, only instance data is serialized. Previous
  instance data is saved to be able to map old versions.
\item[Delta] The object is versioned and buffered. The delta data is typically
  smaller than the instance data. The delta data is transmitted to slave
  instances for synchronization. Previous instance and delta data is saved to be
  able to map and sync old versions.
\item[Unbuffered] The object is versioned and unbuffered. No data is stored, and
  no previous versions can be mapped.
\end{compactdesc}

Serialization is facilitated using output or input streams, which abstract the
data transmission and are used like a \textsf{std::stream}. The data streams
implement efficient buffering and compression, and automatically select the best
connection for data transport. Custom data type serializers can be implemented
by providing the appropriate serialization functions. No pointers should be
directly transmitted through the data streams. For pointers, the corresponding
object is typically a distributed object as well, and its UUID and version are
transmitted in place of a pointer.

Dynamic objects are versioned, and on \textsf{commit} the delta data from the
previous version is sent, if available using multicast, to all mapped slave
instances. The data is queued on the remote node, and is applied when the
application calls \textsf{sync} to synchronize the object to a new version. The
\textsf{sync} method might block if a version has not yet been committed or is
still in transmission. All versioned objects have the following characteristics:

\begin{compactitem}
\item The master instance of the object generates new versions for all
  slaves. These versions are continuous. It is possible to commit on slave
  instances, but special care has to be taken to handle possible
  conflicts.
\item Slave instance versions can only be advanced; that is, \textsf{sync(
  version)} with a version smaller than the current version will fail.
\item Newly mapped slave instances are mapped to the oldest available
  version by default, or to the version specified when calling
  \textsf{mapObject}.
\end{compactitem}

\label{sec:Serializable}The \textsf{Collage} Serializable implements one
convenient usage pattern for object data distribution. The
\textsf{co::Serializable} data distribution is based on the concept of dirty
bits, allowing inheritance with data distribution. Dirty bits form a 64-bit mask
which marks the parts of the object to be distributed during the next commit.
For serialization, the application developer implements \textsf{serialize} or
\textsf{deserialize}, which are called with the bit mask specifying which data
has to be transmitted or received. During a commit or sync, the current dirty
bits are given, whereas during object mapping all dirty bits are passed to the
serialization methods.

Blocking commits allow to limit the number of outstanding, queued versions on
the slave nodes. A token-based protocol will block the commit on the master
instance if too many unsynchronized versions exist.

\subsubsection{Optimizations}

The API presented in the previous section provides sufficient abstraction to
implement various optimizations for faster mapping and synchronization of data:
compression, chunking, caching, preloading and multicast. The results section
evaluates some of these optimizations.

The most obvious one is compression. Recently, many new compression algorithms
have been developed which exploit modern CPU architectures and deliver
compression rates well above one gigabyte per second. \textsf{Collage} uses the
Pression library~\cite{pression}, which provides a unified interface for a
number of compression libraries, such as FastLZ~\cite{jesperfast},
Snappy~\cite{snappy} and ZStandard~\cite{zstd}. It also contains a custom,
virtually zero-cost RLE compressor. Pression parallelizes the compression and
decompression using data decomposition. This compression is generic, and
implemented transparently for the application. Applications can also use
data-specific compression.

The data streaming interface implements chunking, which pipelines the
serialization code with the network transmission. After a configurable number of
bytes has been serialized to the internal buffer, it is transmitted and
serialization continues. This is used both for the initial mapping data and for
commit data.

Caching retains instance data of objects in a client-side cache, and reuses this
data to accelerate mapping of objects. The instance cache is either filled by
``snooping'' on multicast transmissions or by an explicit preloading when the
master objects are registered. The preloading sends instance data of recently
registered master objects to all connected nodes, while the corresponding node
is idle. These nodes simply enter the received data to their cache. Preloading
uses multicast when available.

Due to the master-slave model of data distribution, multicast is used to
optimize the transmission time of data. If the contributing nodes share a
multicast session, and more than one slave instance is mapped, \textsf{Collage}
automatically uses the multicast connection to send the new version information.

\subsection{Barriers, Queues and Object Maps}\label{sec:barrier}

\textsf{Collage} implements a few generic distributed objects which are used by
\textsf{Equalizer} and other applications. A barrier is a distributed barrier
primitive used for software swap synchronization in \textsf{Equalizer}
(\sref{sec:swap}). Its implementation follows a simple master-slave approach,
which has shown to by sufficient for this use case. Queues are distributed,
single producer, multiple consumer FIFO queues. To hide network latencies,
consumers prefetch items into a local queue. Queues are used for tile and chunk
compounds (\sref{sec:tile}).

The object map facilitates distribution and synchronization of a collection of
distributed objects. Master versions can be registered on a central node, e.g.,
the application node in \textsf{Equalizer}. Consumers, e.g., \textsf{Equalizer}
render clients, can selectively map the objects they are interested in.
Committing the object map will commit all registered objects and sync their new
version to the slaves. Syncing the map on the slaves will synchronize all mapped
instances to the new version recorded in the object map. This effective design
allows data distribution with minimal application logic. It is used by
\textsf{Sequel} (\sref{sec:sequel}) and other \textsf{Collage} applications.

\section{Example Applications} \label{sApplications}

In this section, we present some major applications built using
\textsf{Equalizer}, and show how they interact with the framework to solve
complex parallel rendering problems.

\subsection{Livre}

Livre (Large-scale Interactive Volume Rendering Engine) is a GPU ray-casting based parallel
4D volume renderer, implementing state-of-the-art view-dependent level-of-detail rendering (LOD)
and out-of-core data management~\cite{EHKRW:06}. Hierarchical and out-of-core LOD
data management is supported by an implicit volume octree, accessed asynchronously
by the renderer from a data source on a shared file system. Different data sources providing
an octree conform access to RAW or compressed as well as to implicitly generated volume data
(e.g. such as from event simulations or surface meshes) can be used.

High-level state information, e.g., camera position and rendering settings, is shared in Livre
through \textsf{Collage} objects between the parallel applications and rendering threads.
Sort-first decomposition is efficiently supported through octree traversal and culling both
for scalability as well as for driving large-scale tiled display walls.

\subsection{RTT Deltagen}

RTT Deltagen (now Dassault 3D Excite) is a commercial application for
interactive, high quality rendering of CAD data. The RTT Scale module,
delivering multi-GPU and distributed execution, is based on \textsf{Equalizer}
and \textsf{Collage}, and has driven many of the aforementioned features.

RTT Scale uses a master-slave execution mode, were a single running Deltagen
instance can go into ``Scale mode'' at any time by launching an
\textsf{Equalizer} configuration. Consequently, the whole internal
representation needed for rendering is based on a \textsf{Collage}-based data
distribution. The rendering clients are separate, smaller applications which
will map their scenes during startup. At runtime, any change performed in the
main application is committed as a delta at the beginning of the next frame,
following a design pattern similar to the \textsf{Collage Serializable}
(\sref{sec:Serializable}). Multicast (\sref{sec:RSP}) is used to keep data
distribution times during session launch reasonable for larger cluster sizes
(tens to hundreds of nodes).

RTT Scale is used for a wide variety of use cases. In virtual reality, the
application is used for virtual prototyping and design reviews in front of
high-resolution display walls and CAVEs, as well as for virtual prototyping of
human-machine interactions using CAVEs and HMDs (\fig{FIG_teaser}(b)). For
scalability, sort-first and tile compounds are used to achieve fast,
high-quality rendering, primarily for interactive raytracing, both based on CPUs
and GPUs. For CPU-based raytracing, often Linux-based rendering clients are used
with a Windows-based application node.

\subsection{RTNeuron}

RTNeuron \cite{HBBES:13} is a scalable real-time rendering tool for the
visualisation of neuronal simulations based on cable models. It is based on
OpenSceneGraph for data management and Equalizer for parallel rendering, and
focuses not only on fast rendering times, but also on fast loading times with no
offline preprocessing. It provides level of detail (LOD) rendering, high quality
anti-aliasing based on jittered frusta and accumulation during still views,
interactive modification of the visual representation of neurons on a per-neuron
basis (full neuron vs. soma only, branch pruning depending on the branch level,
\dots). RTNeuron implements both sort-first and sort-last rendering with order
independent transparency.

\subsection{RASTeR}

RASTeR~\cite{BGP:09} is an out-of-core and view-dependent real-time multiresolution
terrain rendering approach using a patch-based restricted quadtree triangulation.
For load-balanced parallel rendering~\cite{GMBP:10} it exploits fast hierarchical
view-frustum culling of the level-of-detail (LOD) quadtree for sort-first decomposition, and
uniform distribution of the visible LOD triangle patches for sort-last decomposition.
The latter is enabled by a fast traversal of the patch-based restricted quadtree
triangulation hierarchy which results in a list of selected LOD nodes,
constituting a view-dependent cut or \emph{front of activated nodes} through the LOD hierarchy.
Assigning and distributing equally sized segments of this active LOD front to the concurrent
rendering threads results in a near-optimal sort-last decomposition for each frame.

\subsection{Bino}

Bino is a stereoscopic 3D video player capable of running on very large display
systems. Originally written for the immersive semi-cylindrical projection
system at the University of Siegen, it has been used in many installations
thanks to its flexibility of configuration. Bino decodes the video on each
Equalizer rendering process, and only synchronizes the time step globally,
therefore providing a scalable solution to video playback.

\subsection{Omegalib}

Omegalib \cite{Omegalib} is a software framework built on top of Equalizer that
facilitates application development for hybrid reality environments such as the
Cave 2. Hybrid reality enviroments aim to create a seamless 2D/3D environment
that supports both information-rich analysis (traditionally done on tiled
display wall) as well as virtual reality simulation exploration (traditionally
done in VR systems) at a resolution matching human visual acuity. Omegalib
supports dynamic reconfigurability of the display environment, so that areas of
the display can be interactively allocated to 2D or 3D workspaces as needed. It
makes it possible to have multiple immersive applications running on a
cluster-controlled display system, have different input sources dynamically
routed to applications, and have rendering results optionally redirected to a
distributed compositing manager. Omegalib supports pluggable front-ends, to
simplify the integration of third-party libraries like OpenGL, OpenSceneGraph,
and the Visualization Toolkit (VTK).

\section{Experimental Results}\label{sec:results}

This section presents new experiments, complementing the results of previous
publications~\cite{EP:07, EMP:09, MEP:10, EEP:11, EBAHMP:12, HBBES:13, deflect,
SPEP:16}. The first part summarizes rendering performance over all decomposition
modes with a few representative workloads. The second part analyses data
distribution performance, in particular how the optimizations in
\textsf{Collage} perform in realistic scenarios.

\subsection{Decomposition Modes}


We conducted new performance benchmarks for various decomposition modes on a cluster
using hexacore Intel Xeon E5-2620v3 CPUs (2.4\,GHz), nVidia GTX 970 GPUs with 4\,GB VRAM each, 16\,GB main
memory per node, 4\,GBit/s Ethernet, and QDR Infiniband. GCC 4.8 has been used with CMake
3.7 release mode settings to compile the software stack.

\begin{figure*}[ht]%
\includegraphics[height=.17\textwidth]{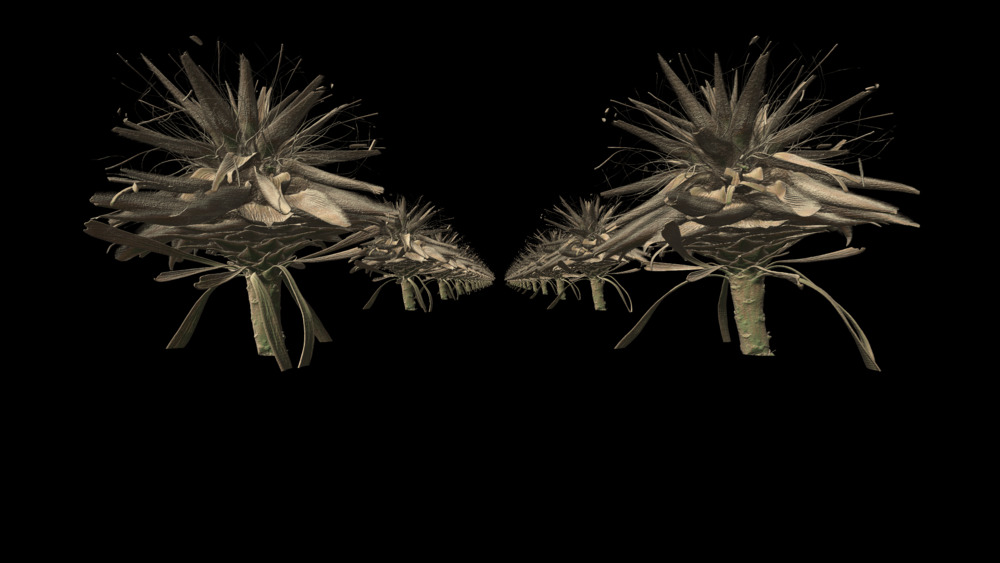}\hfil
\includegraphics[height=.17\textwidth]{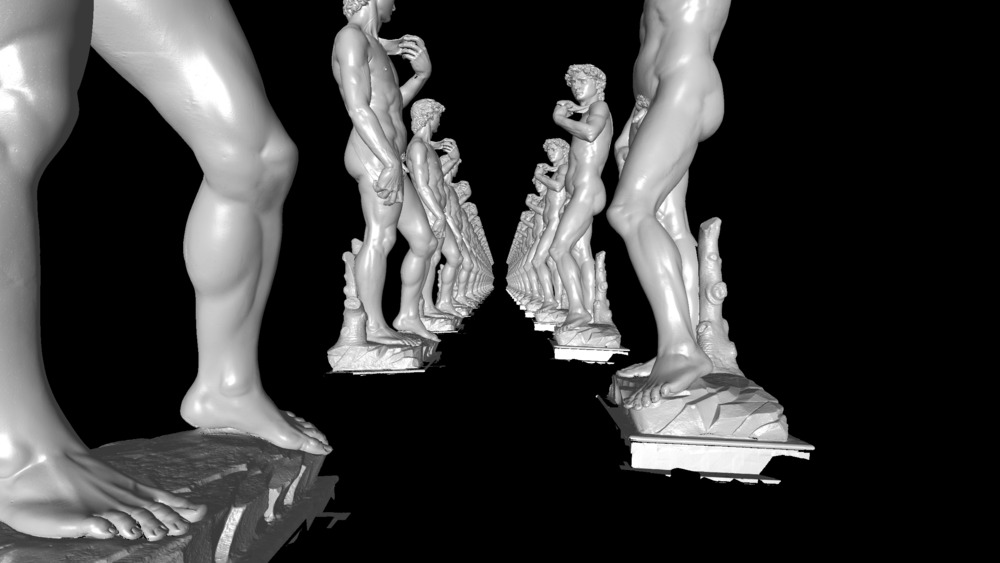}\hfil
\includegraphics[height=.17\textwidth]{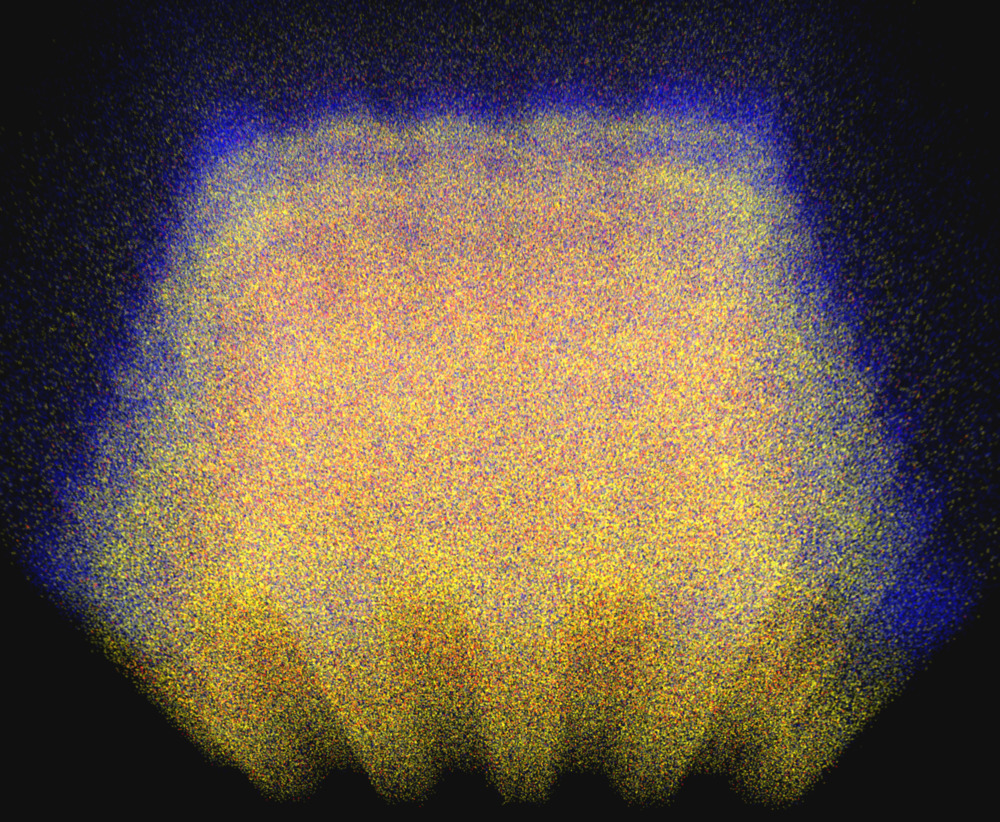}\hspace{.2mm}\includegraphics[height=.17\textwidth]{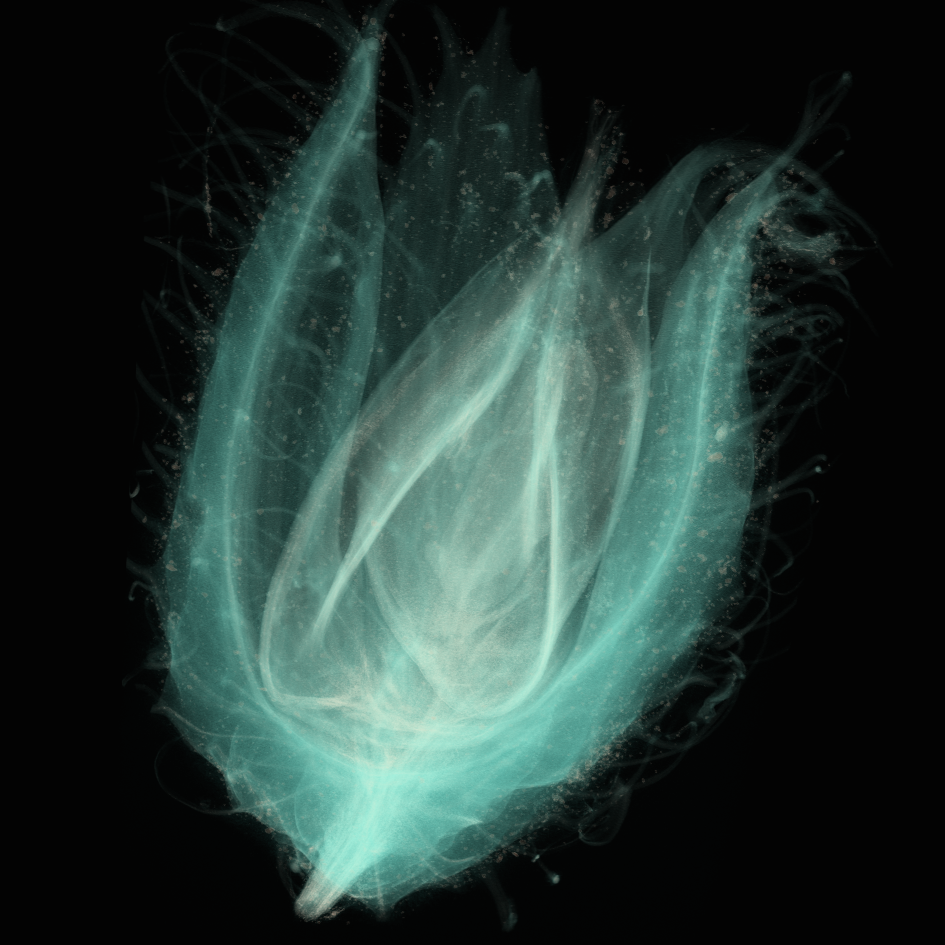}
  \caption{\label{fBenchmarks}Benchmark data: Flower alley with 30 volume models
  of $1024^3$ unsigned bytes (1\,GB) each (left); David alley with 30 statues of
  56\,M triangles (988\,MB) each (middle); Volumes used for object distribution
  benchmarks (right): spike frequencies of a three million neuron electrical
  simulation at $512\times437\times240$ unsigned byte (51\,MB) and an MicroCT scan of a
  beechnut at $1024\times1024\times1546$ unsigned short (3\,GB) resolution}
\end{figure*}

We tested the decomposition modes with both polygonal data and volume data
(\fig{fBenchmarks} (middle and left)), using test scenes that allowed to adapt
the rendering load the system has to cope with. In both cases the scene is
comprised of two rows of instantiated, identical models with 30 models in total.
Rendering was performed at an output resolution of $2560\times1440$. The camera
is initially placed in the center of the scene, between the two rows, rendering
only half of the model instances. It is then moved backward over the duration of
800 frames, steadily increasing the rendering load by revealing more models,
until all 30 instances are visible.

\begin{figure*}[ht]\center
  \includegraphics[width=\textwidth]{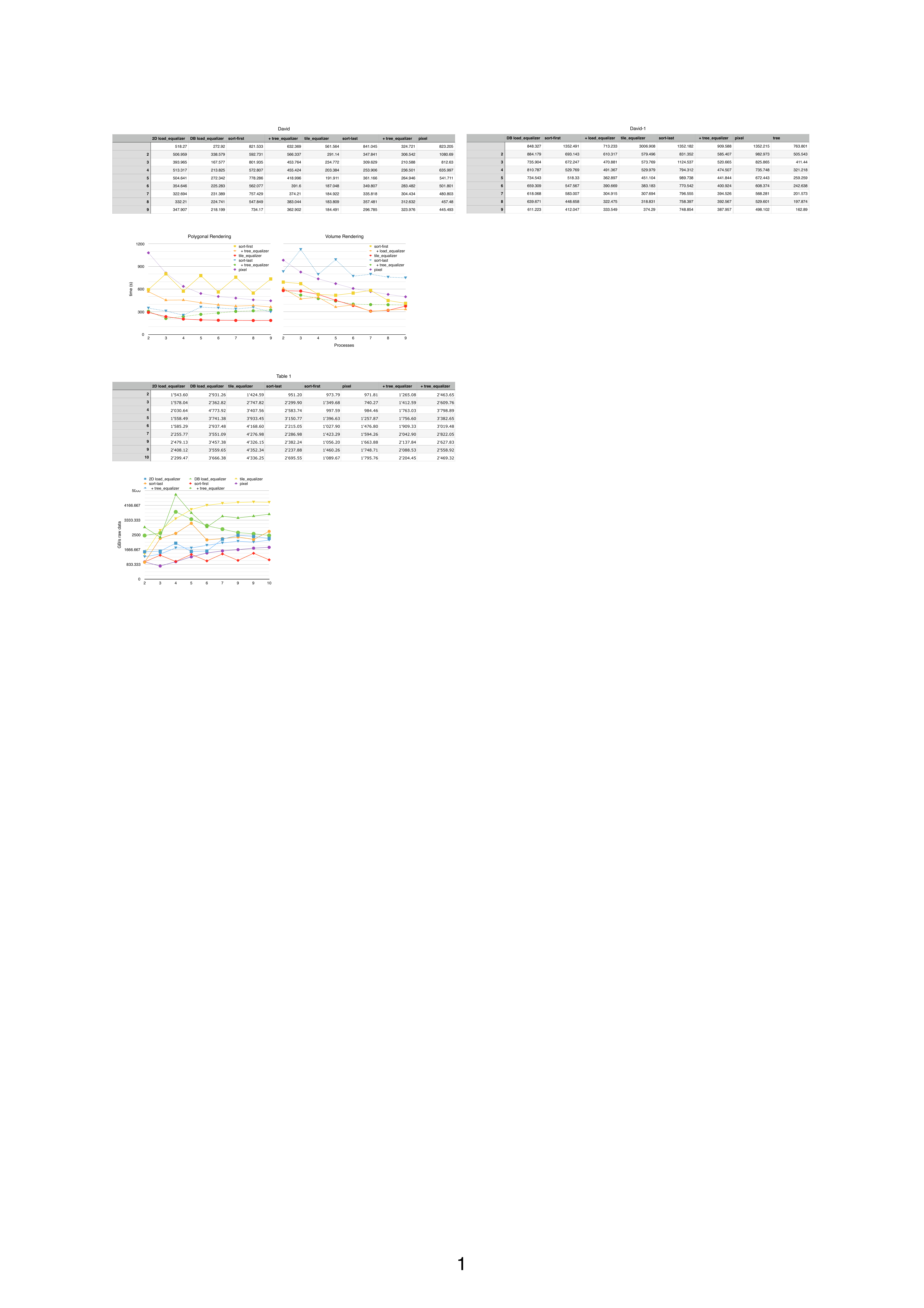}
  \caption{\label{fCompounds}Compound scalability for polygonal and volume data}
\end{figure*}

We investigated the scalability of individual decomposition modes by running the
same experiment using a varying number of render nodes (2-9) and one dedicated
application/display node. We subsequently summed up the duration of all rendered
frames for each run (\fig{fCompounds}).

For sort-first and sort-last rendering we present static and load-balanced task
decomposition. For readability, we only present the results of the equalizer
(load or tree) providing the better performance for each application.
Unsurprisingly, static decompositions perform worse over load-balanced
compounds. Sort-first polygon rendering exhibits oscillations in performance as
nodes are added to the task, due to unfavorable assignment of a tile with a high
work load on odd node counts. Static sort-last volume rendering has a similar
oscillation behaviour, as ranges of scene geometry tend to also get unfavorably
assigned under such conditions.

The simpler \textsf{tree\_equalizer} outperforms in almost all cases the
load-grid-driven \textsf{load\_equalizer}, except for sort-first volume
rendering where the load in the region of interest is relatively uniform. This
counterintuitive result seems to again confirm that simple algorithms often
outperform theoretically better, but more complex implementations. On the other
hand, the \textsf{tile\_equalizer} often outperforms \textsf{tree\_equalizer}.
This suggests that the underlying implicit load balancing can be superior to the
explicit methods of \textsf{load\_equalizer} and \textsf{tree\_equalizer} in
high load situations, where the additional overhead of tile generation and
distribution is more justified. The relatively simple nature of our benchmark
application's rendering algorithms is also favoring work packages, since they
have a near-zero static overhead per rendering pass.

Finally, we also provide scalability results for pixel compounds. While
naturally load-balanced, pixel compounds only scale fill rate and not geometry
processing. Consequently, pixel compounds provide better performance for volume
rendering, and a predictable scaling behaviour for both.

\begin{figure*}[ht]\center
  \includegraphics[width=\textwidth]{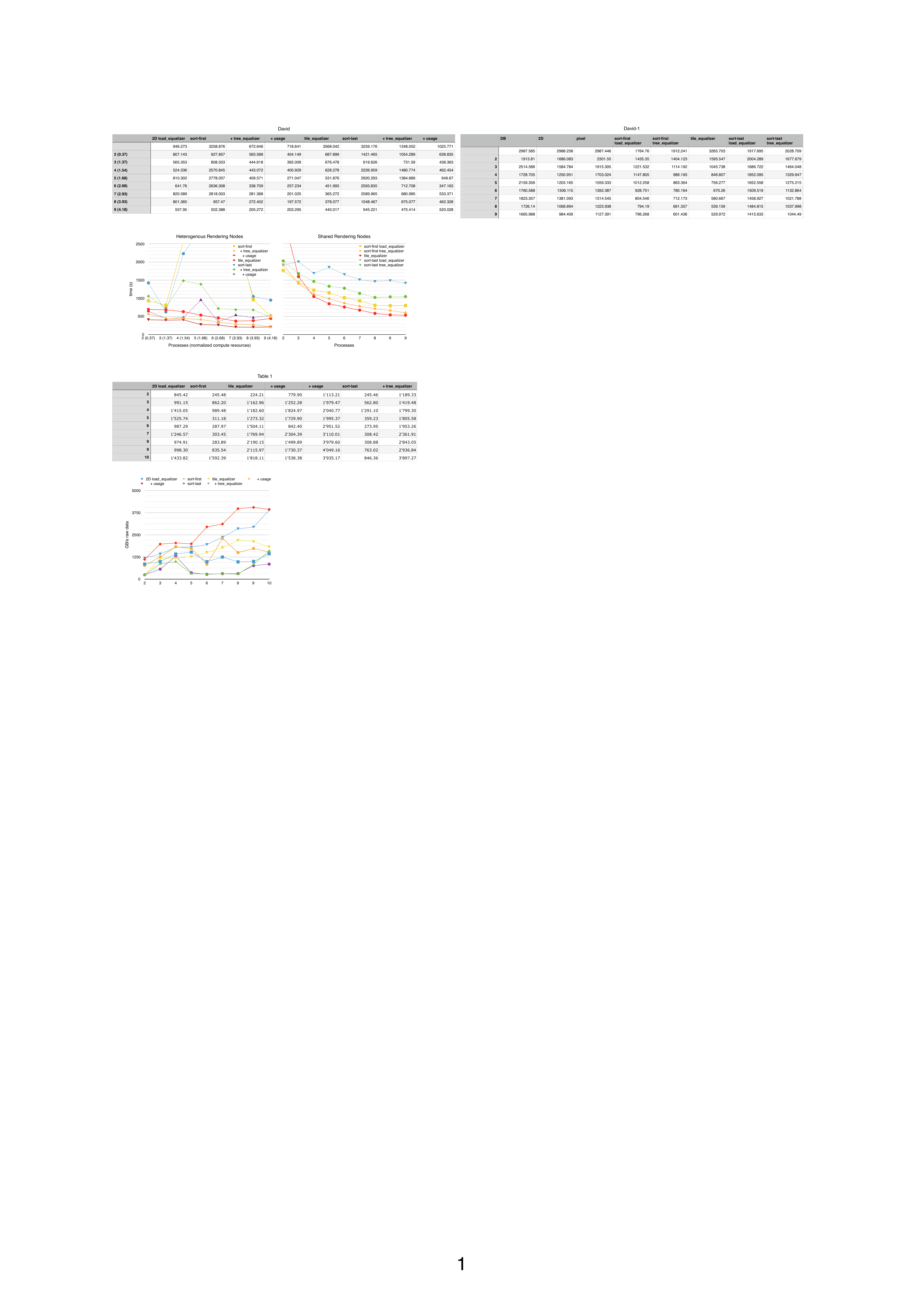}
  \caption{\label{fEqualizers}Scalability with heterogenous rendering resources}
\end{figure*}

For volume rendering we also measured the performance of decomposition modes
under heterogeneous load, which was easily achievable by varying the number of
volume samples used for each fragment (1-7) while rendering. This allowed for a
consistent linear scaling of rendering load, which was randomly varied
(\fig{fEqualizers}) either per frame or per node. Such a linear scaling of load
per node corresponds to a scaling of resources, e.g., doubling the rendering
load on a specific node corresponds to halving its available rendering
resources. To the system this node would then contribute the value 0.5 in terms
of {\em normalized compute resources}, as illustrated by \fig{fEqualizers}
(left).

This figure gives an impression of how individual modes perform on heterogeneous
systems. In this case the tree equalizer performs best (\fig{fEqualizers}
(left)), as it allows us to a priori define how much {\em usage} it should make
of individual nodes, i.e., bias the allocation of rendering time in accordance
with the (simulated) compute resources. \fig{fEqualizers} (right), on the other
hand, illustrates how the tested decomposition modes perform on a system where
compute resources fluctuate randomly every frame, as can arguably be the case
for shared rendering nodes in virtualized enviroments. For this scenario tile
equalizer seems best suited, as it performs load balancing implicitly and does
not assume coherence of available resources between frames. The simpler tree
equalizer also outperforms the load equalizer in this experiment.

\subsection{Object Distribution}

The data distribution benchmarks have been performed on a cluster using dual
processor Intel Xeon X5690 CPUs (3.47\,GHz), 24\,GB main memory per node,
10\,GBit/s Ethernet and QDR Infiniband. Intel ICC 2017 has been used with CMake
3.2 release mode settings to compile the software stack. To benchmark the data
distribution we used two datasets: The David statue at 2\,mm resolution (as in
\fig{fBenchmarks} (middle), but in 2\,mm resolution) and 3D volumes of spike
frequencies of an electrical simulation of three million neurons
(\fig{fBenchmarks} (right)).

The PLY file is converted into a k-d tree for fast view frustum culling and
rendering, and the resulting data structure is serialized in binary form for
data transmission. The spike frequency volumes aggregate the number of spikes
which happened within a given voxel over a given time. The absolute spike count
is renormalized to an unsigned byte value during creation. Higher densities in
the volume represent higher spiking activity in the voxel.

\subsubsection{Data Compression Engines}

\begin{figure}[ht]\center
  \includegraphics[width=\columnwidth]{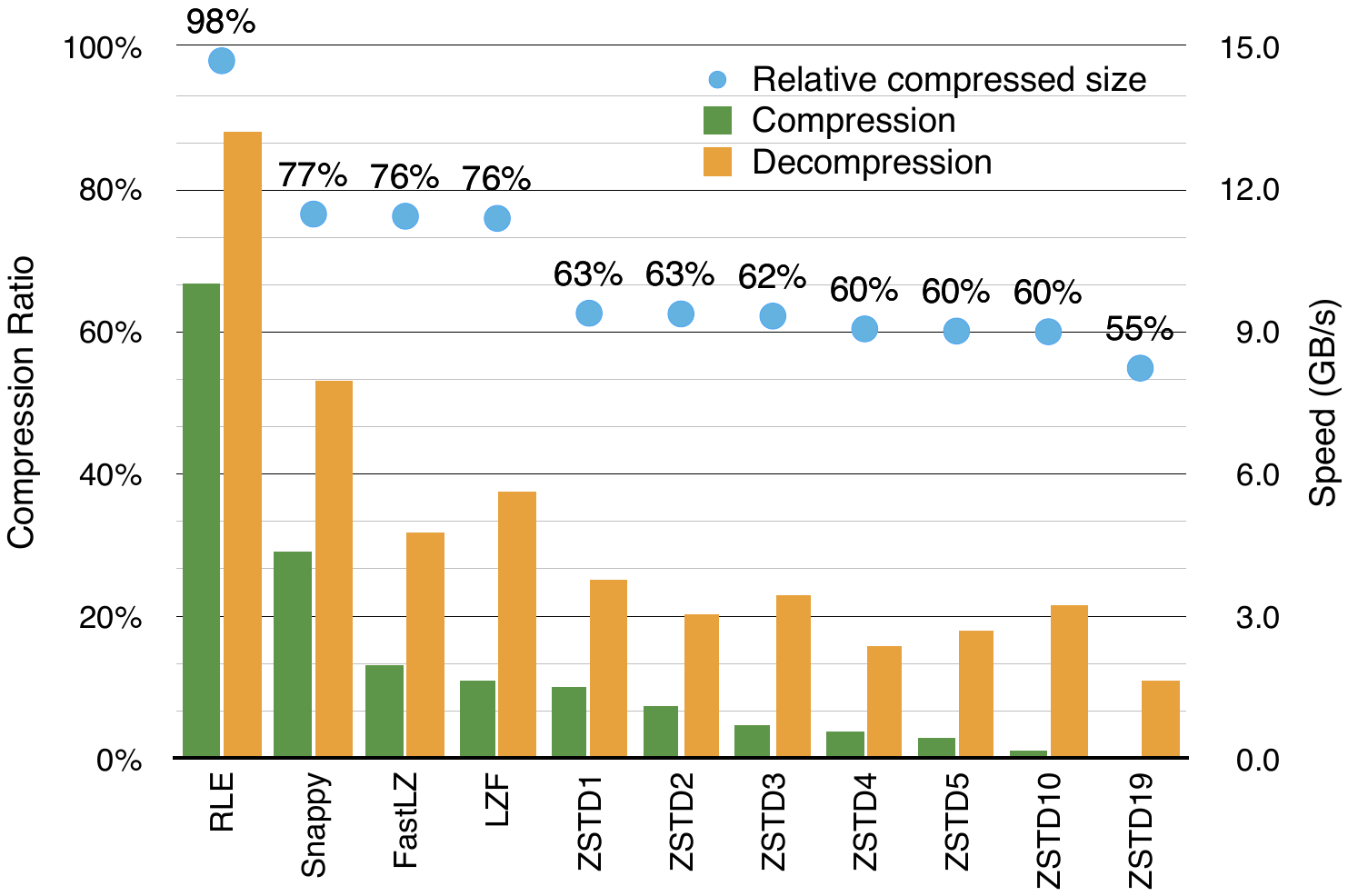}
  \caption{\label{fCompressor}Data compression for generic binary data}
\end{figure}

A critical piece for data distribution performance are the characteristics of
the data compression algorithm. Our microbenchmark compresses a set of binary
files, precalculating the speed and compression ratio of the various engines.
\fig{fCompressor} shows the compression and decompression speed in gigabyte per
second as well as the size of the compressed data relative to the uncompressed
data. The ZSTD$x$ engines use the ZStandard compression library at compression
level $x$. The measurements were performed on a single, isolated node.

RLE compression has a very low overhead but merely removes ``blank space'' in
the data. The Snappy compression, used as default in \textsf{Collage}, achieves
the same compression ratio as the LZ variants at a much higher speed. The
ZStandard compressor has roughly the same speed as the LZ variants at the lowest
compression level, but provides significantly better compression. At higher
compression levels it can improve the compression ratio slightly, but at a high
cost for the compression speed.

The compression ratio for the models used in the following section deviate from
this averaged distribution. \fig{fCompressorDetail} shows the compression ratios
for the triangle and volume data.

The PLY data is little compressible, the default compressor achieves a 10\%
reduction. This is due to a high entropy in the data and the dominant use of
floating point values. Overall, the profile is similar to the generic benchmark,
at a smaller compression rate.

The volume data on the other hand is sparsely populated and using integer (byte
and short) values, which is easier to compress. The naive RLE implementation
already achieves a good compression rate, showing that the smaller volume contains at
most 28\% empty space and the bigger volume at most 43\%. Snappy and ZStandard
can reduce the spike data much further, reducing the data to a few megabytes.
Surprisingly, the beechnut data set does not yield significantly higher
compression with the modern Snappy and ZStandard libraries.

\subsubsection{Model Distribution and Update}

In this section we analyse how data distribution and synchronization performs in
real-world applications. We extracted the existing data distribution code from a
mesh renderer (eqPly) and a volume renderer (Livre) into a benchmark application
to measure the time to initially map all the objects on the render client nodes,
and to perform a commit-sync of the full data set after mapping has been
established. All figures observe a noticable measurement jitter due to other
services running on the cluster. The details of the benchmark algorithm can be
found in the
\htmladdnormallinkfoot{implementation}{https://github.com/Eyescale/Equalizer/tree/paper2018/tools/eqObjectBench}.

We used the same data sets as in the previous section, and ran the benchmark on up
to eight physical nodes, that is, after eight processes nodes start to run two
processes per node, which share CPU, memory and network interface bandwidth.

Object mapping is measured using the following settings: \textbf{none}
distributes the raw, uncompressed, and unbuffered data, \textbf{compression}
uses the Snappy compressor to distribute unbuffered data, \textbf{buffered}
reuses uncompressed, serialized data for mappings from multiple nodes, and
\textbf{compression buffered} reused the compressed buffer for multiple nodes.
Unbuffered operations need to reserialize, and potentially recompress, the
master object data for each slave node. Each slave instance needs to deserialize
and decompress the data.

During data synchronization, the master commits the object data to all mapped
slave instances simultaneously. This is a ``push'' operation, whereas the
mapping is a slave-triggered ``pull'' operation. Slave nodes queue this data and
consume it during synchronization. We test the time to commit and sync the data
using different compression engines.

\begin{figure*}[p!]\center
  \includegraphics[height=.19\textheight]{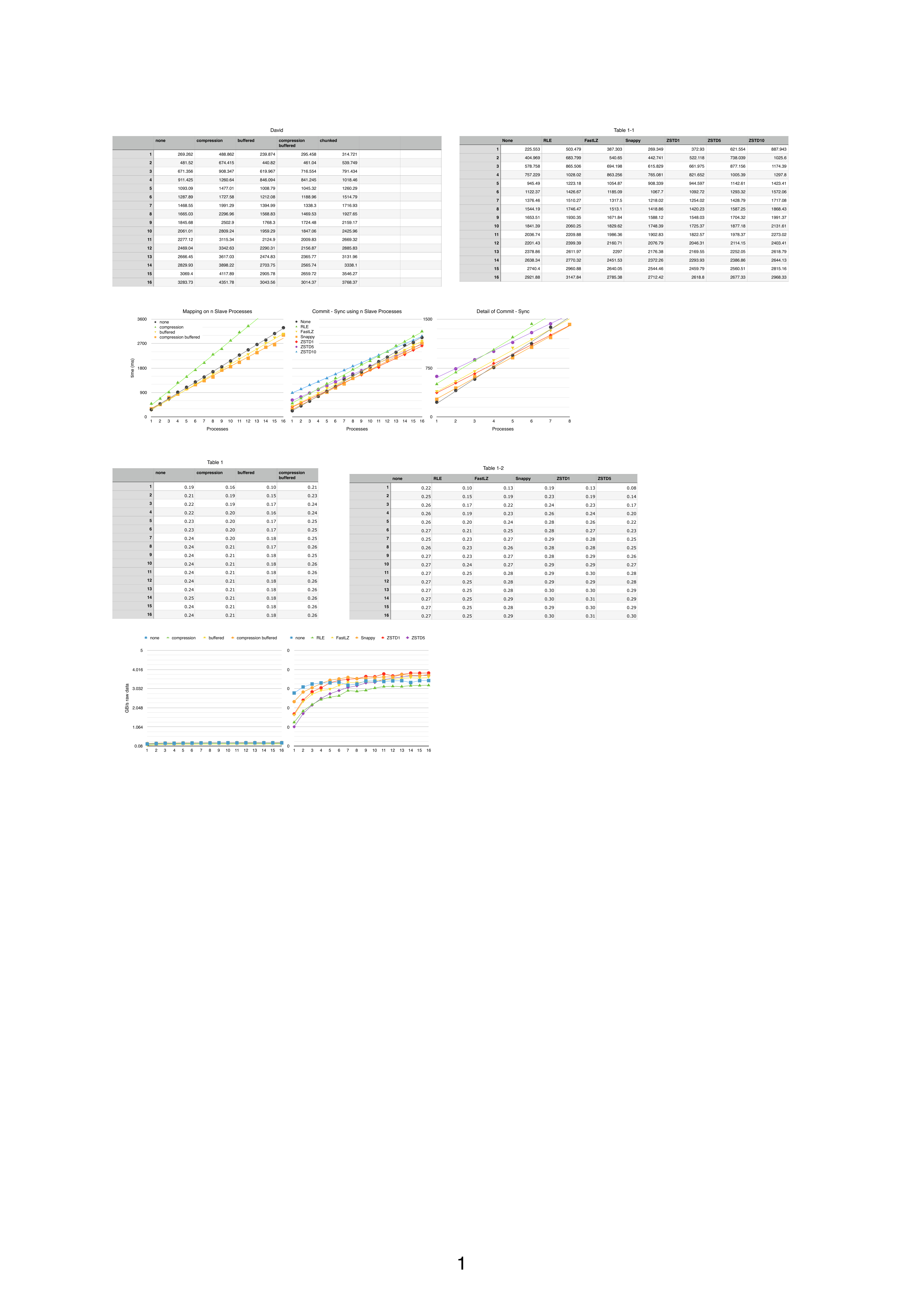}
  \caption{\label{fDavidDist}Object mapping (left) and data synchronization
    time (middle, detail view right) for the David 2\,mm data set}
\end{figure*}

The David statue at 2\,mm resolution is organized in a k-d tree for rendering. Each
node is a separate distributed object, which has two child node objects. A
total of $1023$ objects is distributed and synchronized. \fig{fDavidDist} shows
the distribution times for this data set. Due to limited compressibility of the
data, the results are relatively similar. Compressing the data repeatedly for
each client leads to decreased performance, since the compression overhead
cannot be amortized by the decreased transmission time. Buffering data slightly
improves performance by reducing the CPU and copy overhead. Combining
compression and buffering leads to the best performance, although only by about
10\%.

During synchronization, data is pushed from the master process to all mapped
slaves using a unicast connection to each slave. While the results in the
\fig{fDavidDist} (middle) are relatively close to each other, we can still
observe how the tradeoff between compression ratio and speed influences overall
performance. Better, slower compression algorithms lead to improved overall
performance when amortized over many send operations.

\begin{figure*}[p!]\center
  \includegraphics[height=.19\textheight]{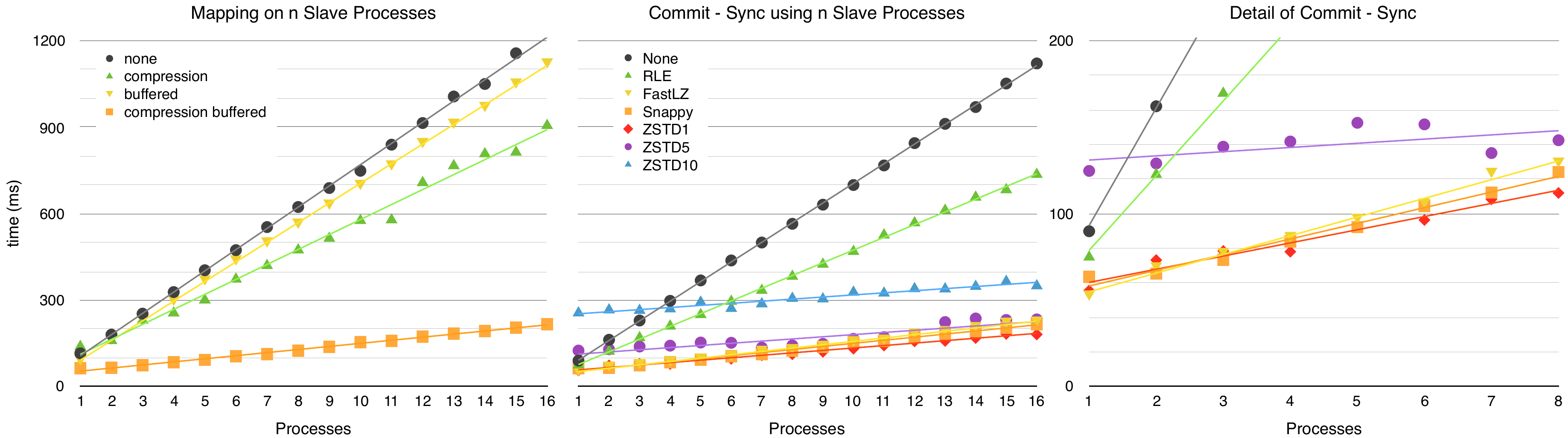}
  \caption{\label{fSpikesDist}Object mapping (left) and data synchronization
    time (middle, detail view right) for the Spike data set in
    \fig{fBenchmarks} (right)}
\end{figure*}

The volume data sets are distributed in a single object, serializing the raw
volume buffer. The Spike volume data set has a significant compression ratio,
which is reflected by the results in \fig{fSpikesDist}. Compression for this
data is beneficial for transmitting data over a 10\,Gb/s link even for a single
slave process. Buffering has little benefit since the serialization of volume
data is trivial. Buffered compression makes a significant difference, since the
compression cost can be amortized over $n$ nodes, reaching raw data transmission
rates of 3.7\,GB/s with the default Snappy compressor and at best 4.4\,GB/s with
ZStandard at level 1.

\begin{figure*}[ht]\center
  \includegraphics[width=\textwidth]{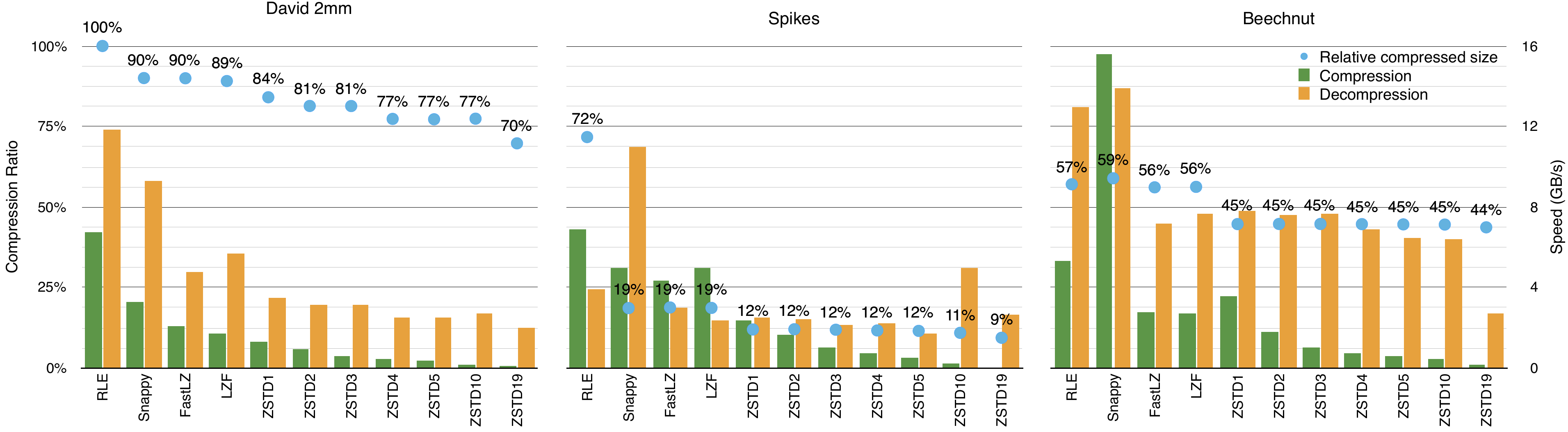}
  \caption{\label{fCompressorDetail}Data compression for PLY data (left, David
    statue 2\,mm) and raw volumes shown in \fig{fBenchmarks} (right)}
\end{figure*}

\begin{figure*}[p!]\center
  \includegraphics[height=.19\textheight]{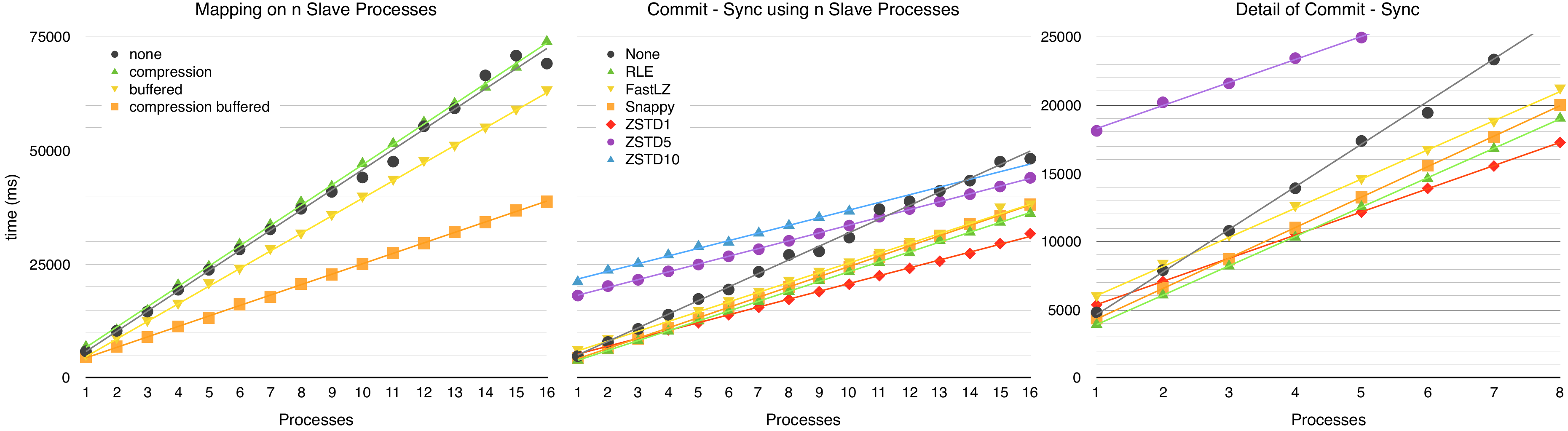}
  \caption{\label{fBeechnutDist}Object mapping (left) and data synchronization
    time (middle, detail view right) for the beechnut data set in
    \fig{fBenchmarks} (right)}
\end{figure*}

The distribution of the beechnut data set also behaves as expected
(\fig{fBeechnutDist}). Due to the larger object size, uncompressed transmission
is slightly faster compared to the Spike data set at 700\,MB/s, and compressed
transmission does not improve the mapping performance, likely due to increased
memory pressure caused by the large data size. The comparison of the various
compression engines is consistent with the benchmarks in
\fig{fCompressorDetail}; RLE, Snappy and the LZ variants are very close to each
other, and ZSTD1 can provide better performance after four nodes due to the
better compression ratio.

\begin{figure*}[p!]\center
  \includegraphics[height=.19\textheight]{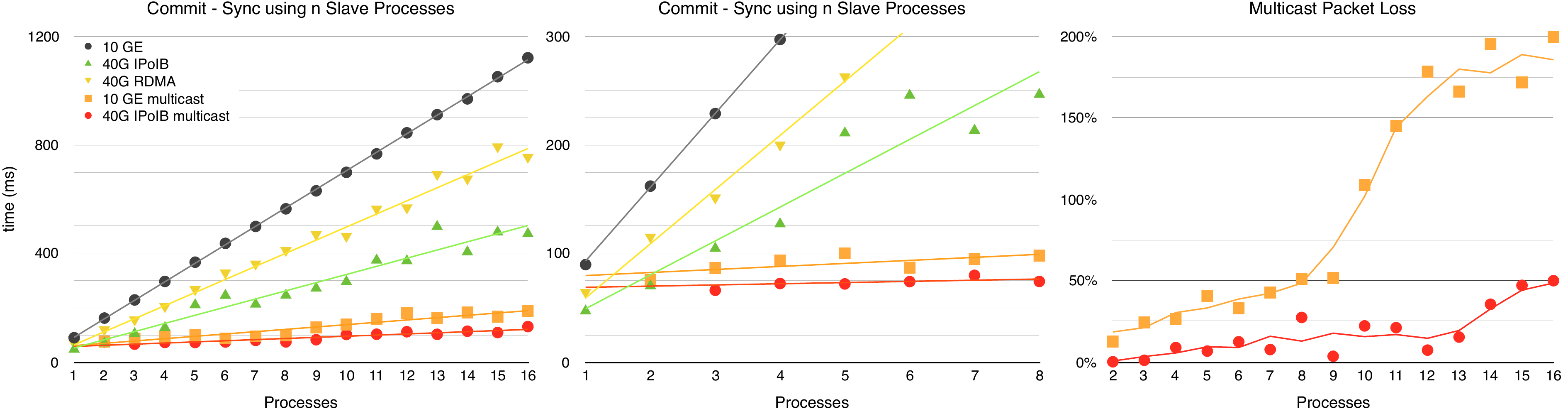}
  \caption{\label{fIFDist}Object synchronization using different network
    transports}
\end{figure*}

Finally, we compare data distribution speed using different protocols. In this
benchmark, data synchronization time of the Spike volume data set is measured,
as in \fig{fSpikesDist} (middle). Buffering is enabled, and compression is
disabled to focus on the raw network performance. \fig{fIFDist} shows the
performance using various protocols. TCP over the faster InfiniBand link
outperforms the cheaper ten gigabit ethernet link by more than a factor of two.
Unexpectedly, the native RDMA connection performs worse, even though it
outperforms IPoIB in a simple peer-to-peer connection benchmark. This needs
further investigation, but we suspect the abstraction of a byte stream
connection chosen by \textsf{Collage} is not well suited for remote DMA
semantics; that is, one needs to design the network API around zero-copy
semantics with managed memory for modern high-speed transports. Both Infiniband
connections show significant measurement jitter.

RSP multicast performs as expected. \textsf{Collage} starts using multicast to
commit new object versions when two or more clients are mapped, since the
transmission to a single client is faster using unicast. RSP consistently
outperforms unicast on the same physical interface and shows good scaling
behavior (2.5 times slower on 16 vs 2 clients on ethernet, 1.8 times slower on
InfiniBand). The scaling is significantly better when only one process per node
is used (\fig{fIFDist}, middle: 30\% slower on ethernet, nearly flat on
InfiniBand). The increased transmission time with multiple clients is caused by
a higher probability of packet loss, which increases significantly when using
more than one process per node. \fig{fIFDist} (right) plots the number of
retransmissions divided by the number of datagrams. Infiniband outperforms
ethernet slightly, but is largely limited by the RSP implementation throughput
of preparing and queueing the datagrams to and from the protocol thread, which
we observed in profiling.

\section{Discussion and Conclusion}
\label{sec:conclusions}

We have presented a significantly improved generic parallel rendering system
over the original publication \cite{EMP:09}. While the original publication
motivated the system design, this publication describes a feature-rich, mature
implementation capable of supporting a wide variety of use cases. We doubled the
support for scalable rendering modes, many of which are presented here for the
first time. We present new runtime adaptations for better load balance and
performance, describe how common optimizations are integrated into the system,
making Equalizer the most generically available scalable rendering system.

Furthermore, we present many new features needed in parallel rendering
applications, from advanced Virtual Reality support to advanced display system
setup for 2D/3D integration, auto-configuration and runtime reconfiguration, and
an advanced network data synchronization library tailored to parallel rendering
applications. We highlight a few commercial and research applications
underlining the generic and versatile system implementation.

With respect to the feature set implemented, we believe that Equalizer now
covers almost any scenario within its scope. For future work, we would like to
integrate new research for better scalability, new network implementations in
particular for modern zero-copy RDMA based transports, as well as extending the
Sequel abstraction layer for ease of use.

\appendices
\ifCLASSOPTIONcompsoc
  \section*{Acknowledgments}
\else
  \section*{Acknowledgment}
\fi
We would like to thank and acknowledge the following institutions and projects
for providing the 3D geometry and volume test data sets: the Digital
Michelangelo Project, Stanford 3D Scanning Repository, Cyberware Inc.,
volvis.org and the Visual Human Project.
This work was partially supported by the Swiss National Science Foundation
Grants 200021-116329 and 200020-129525, the Swiss Commission for Technology and
Innovation CTI/KTI Project 9394.2 PFES-ES, the EU FP7 People Programme (Marie
Curie Actions) under REA Grant Agreement n$^{\circ}290227$ and a Hasler Stiftung
grant (project number $12097$).

We would also like to thank all supporters and contributors of
\textsf{Equalizer}, most notably RTT, the Blue Brain Project, the University of
Siegen, the Electronic Visualization Lab at the University of Illinois Chicago
and Dardo Kleiner.


%

\begin{IEEEbiography}[{
\includegraphics[width=1in,height=1.25in,clip,keepaspectratio]{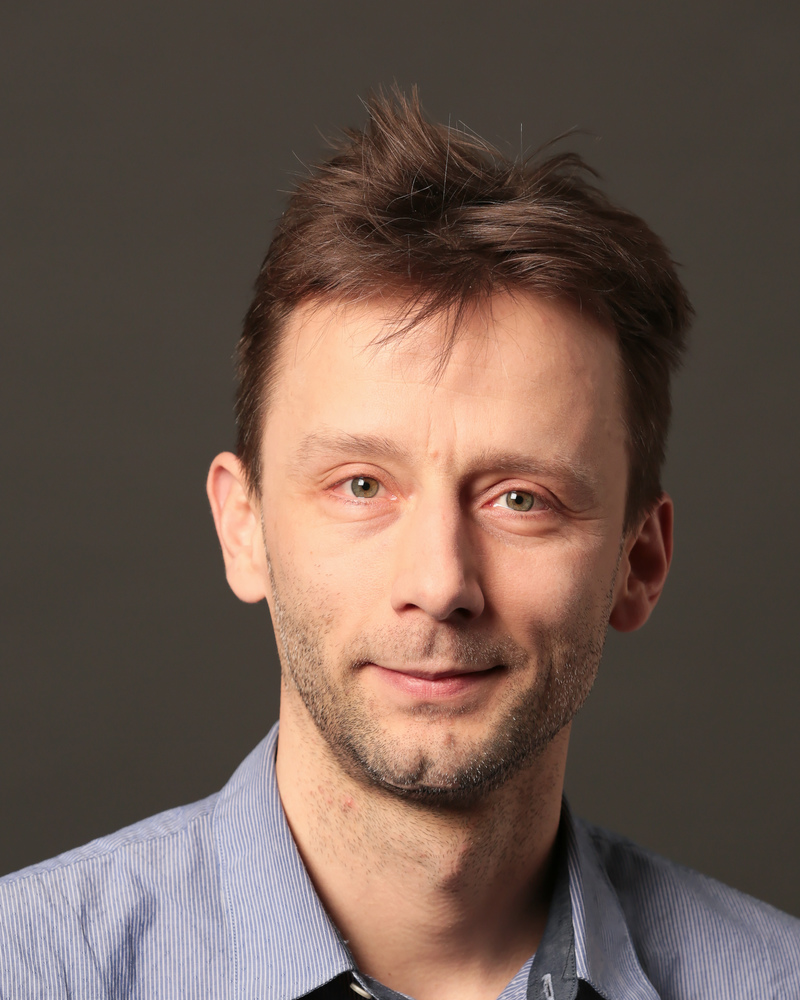}}]{
Stefan Eilemann} works towards large-scale visualization for Exascale
simulations; interactive integration of simulation, analysis and visualization
as well as flexible frameworks for data sharing and dynamic allocation of
heterogenous resources. He was the technical manager of the Visualization Team
in the Blue Brain Project, is the CEO and founder of Eyescale, and the lead
developer of the \textsf{Equalizer} parallel rendering framework. He received
his masters degree in Computer Science from EPFL in 2015, and his Engineering
Diploma in Computer Science in 1998. He is currently working towards a PhD in
Computer Science at the Visualization and MultiMedia Lab at the University of
Zurich. \end{IEEEbiography}

\begin{IEEEbiography}[{\includegraphics[width=1in,height=1.25in,clip,keepaspectratio]{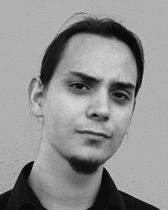}}]{David Steiner}
received MSc degrees from the University of Applied Sciences Upper Austria (Digital Media) and the University of Zurich (Computer Science). He joined the Visualization and MultiMedia Lab (VMML) in 2012 and is currently pursuing his doctorate. His research interests include interactive large-scale data visualization, distributed parallel rendering, and load balancing.
\end{IEEEbiography}

\begin{IEEEbiography}[{\includegraphics[width=1in,height=1.25in,clip,keepaspectratio]{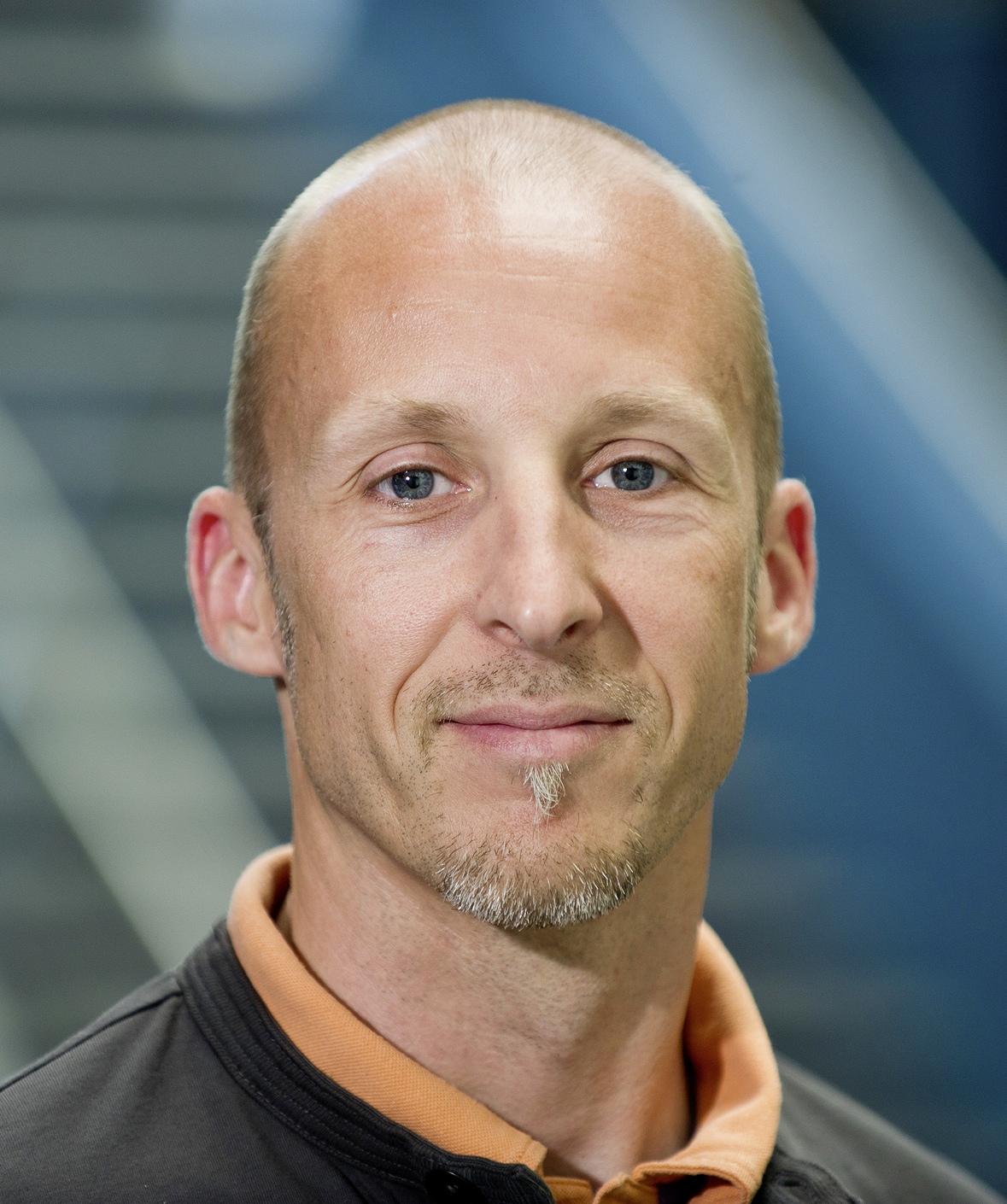}}]{Prof. Dr. Renato Pajarola}
has been a Professor in computer science at the University of Zurich since 2005, leading the Visualization and MultiMedia Lab (VMML). He has previously been an Assistant Professor at the University of California Irvine and a Postdoc at Georgia Tech. He has received his Dipl. Inf-Ing. ETH and Dr. sc. techn. degrees in computer science from the Swiss Federal Institute of Technology (ETH) Zurich in 1994 and 1998 respectively. He is a Fellow of the Eurographics Association and a Senior Member of both ACM and IEEE. His research interests include real-time 3D graphics, interactive data visualization and geometry processing.
\end{IEEEbiography}


\bibliographystyle{abbrv}
\bibliography{references}


\end{document}